\documentstyle[prd,aps,floats]{revtex}

\def\ds{\displaystyle}

\begin{document}

\draft
\input epsf
\twocolumn[\hsize\textwidth\columnwidth\hsize\csname
@twocolumnfalse\endcsname

\title{Q-ball formation: Obstacle to Affleck-Dine baryogenesis \\
in the gauge-mediated SUSY breaking ?}

\author{S. Kasuya and M. Kawasaki}
\address{Research Center for the Early Universe, University 
  of Tokyo, Bunkyo-ku, Tokyo 113-0033, Japan}

\date{June 11, 2001}

\maketitle

\begin{abstract}
We consider the Affleck-Dine baryogenesis comprehensively in the
minimal supersymmetric standard model with gauge-mediated
supersymmetry breaking. Considering the high temperature effects, we
see that the Affleck-Dine field is naturally deformed into the form
of the Q ball. In the natural scenario where the initial amplitude of
the field and the A-terms are both determined by the nonrenormalizable
superpotential, we obtain only very a narrow allowed region in the
parameter space in order to explain the baryon number of the universe
for the case that the Q-ball formation occurs just after 
baryon number production. Moreover, most of the parameter sets suited
have already been excluded by current experiments. We also find 
new situations in which the Q-ball formation takes place rather late
compared with baryon number creation. This situation is more
preferable, since it allows a wider parameter region for naturally
consistent scenarios, although it is still difficult to realize in the
actual cosmological scenario.  
\end{abstract}

\pacs{PACS numbers: 98.80.Cq, 11.27.+d, 11.30.Fs
      \hspace{5cm} hep-ph/0106119}

% 98.80.Cq: Particle-theory and field-theory models of the early
%           Universe (including cosmic pancakes, cosmic strings, 
%           chaotic phenomena, inflationary universe, etc.)
% 11.27.+d: Extended classical solutions; cosmic strings, 
%           domain walls, texture 
% 11.30.Fs: Global symmetries (e.g., baryon number, lepton number)

%\newpage
\vskip2pc]

\setcounter{footnote}{1}
\renewcommand{\thefootnote}{\fnsymbol{footnote}}

\section{Introduction}
The Affleck-Dine (AD) mechanism \cite{AfDi} is the most promising
scenario for explaining the baryon number of the universe. It is based
on the dynamics of a (complex) scalar field $\phi$ carrying baryon 
number, which is called the AD field. During inflation, the
expectation value of the AD field develops at a very large value. 
After inflation the inflaton field oscillates about the minimum of the
effective potential, and dominates the energy density of the universe
like matter, while the AD field stays at the large field value. It 
starts the oscillation, or more precisely, rotation in its effective
potential when $H \sim m_{\phi,eff}$, where $H$ and 
$m_{\phi,eff}\equiv|V''(\phi)|$ are the
Hubble parameter and the curvature of the potential of the AD
field. Once it rotates, the baryon number will be created as 
$n_{B} \sim \omega \phi^2$, where $\omega = \dot{\theta}$ is the
velocity of the phase of the AD field. When $t \sim \Gamma_{\phi}$,
where $\Gamma_{\phi}$ is the decay rate of the AD field, it decays
into ordinary particles carrying baryon number such as quarks, and the 
baryogenesis in the universe completes.

However, important effects on the field dynamics were overlooked. It
was recently revealed that the AD field feels spatial instabilities
\cite{KuSh}. Those instabilities grow very large and the AD field
deforms into clumpy objects: Q balls. A Q ball is a kind of
nontopological soliton, whose stability is guaranteed by the existence
of some charge $Q$ \cite{Coleman}. In the previous work \cite{KK1}, we
found that all the charges which are carried by the AD field are
absorbed into formed Q balls, and this implies that the baryon number
of the universe cannot be explained by the relic AD field remaining
outside Q balls after their formation.

In the radiation dominated universe, charges are evaporated from Q
balls \cite{LaSh}, and they will explain the baryon number of the
universe \cite{KK3}. This is because the minimum of the (free) energy
is achieved when the AD particles are freely in the thermal plasma at
finite temperature. (Of course, the mixture of the Q-ball
configuration and free particles is the minimum of the free
energy at finite temperature, when the chemical potential of the Q
ball and the plasma are equal to be in the chemical equilibrium. This
situation can be achieved for very large charge of Q balls with more
than $10^{40}$ or so \cite{LaSh}.) Even if the radiation component is
not dominant energy in the universe, such as that during the
inflaton-oscillation dominant stage just after the inflation, high
temperature effects on the dynamics of the AD field and/or Q-ball
evaporation are important. Therefore, in this article, we investigate
the whole scenario of the AD mechanism for baryogenesis in the minimal
supersymmetric standard model (MSSM) with the gauge-mediated
supersymmetry (SUSY) breaking in the high temperature universe.  

In Sec.II, we identify one of the flat directions as the AD field,
and look for the form of the effective potential in the context. We
will note how the AD field produces the baryon number in the universe
in Sec.III. In Sec.IV, we consider the dynamics of the (linearized)
fluctuations. The Q-ball formations are simulated in Sec.V, and we
will construct the formula of the Q-ball charge in terms of the
initial amplitude of the AD field. Section VI has to do with the
mechanism for the evaporation of the Q-ball charge, and we estimate
the total charge evaporated, which will be the baryons in the universe
later. In Sec.VII, we seek for the consistent cosmological Q-ball
scenario, and find it very difficult as contrary to our expectation.
In Sec.VIII, we will investigate the Q-ball formation and natural
scenario in more generic gauge-mediation model, where we take the
different scales for the potential height and the messenger scale of
gauge-mediation. In Sec.IX, we will consider a new situation where
the baryon number creation takes place when the field starts the
rotation, but the Q-ball formation occurs rather later. In that case,
we will find the wider consistent regions of the parameter space in
some situations. The detections of the dark matter Q ball and their
constraints are described in Sec.IX. Section X is devoted for our
conclusions.

\section{Flat directions as the Affleck-Dine field}
A flat direction is the direction in which the effective potential
vanishes. There are many flat directions in the minimal supersymmetric 
standard model (MSSM), and they are listed in
Refs.~\cite{DiRaTh,GhKoMa}. Since they consist of squarks and/or
sleptons, they carry baryon and/or lepton numbers, and we can identify 
them as the Affleck-Dine (AD) field. Although the flat directions are
exactly flat when supersymmetry (SUSY) is unbroken, it will be lifted
by SUSY breaking effects. In the gauge-mediated SUSY breaking
scenario, SUSY breaking effects appear at low energy scales, so the
shape of the effective potential for the flat direction has curvature
of order of the electroweak mass at low scales, and almost flat at
larger scales (actually, it grows logarithmically as the field becomes 
larger). To be concrete, we adopt the following special form to
represent such a kind of the potential \cite{KuSh}:
\begin{equation}
    \label{pot-1}
    V(\Phi) = m_{\phi}^4 \log 
                  \left( 1+\frac{|\Phi|^2}{m_{\phi}^2} \right),
\end{equation}
where $\Phi$ is the complex scalar field representing the flat
direction. In spite of the specific form, it includes all the
important features for the formation of Q balls, since they are formed
at large field value, where the logarithmic function has to do with
the dynamics of the field. \footnote{
  In Ref.~\cite{DeMoMu}, the effective potential has the form
  \[
      \nonumber
      V \sim F^2 [\log(\phi^2)]^2, 
  \]
  where $F^{1/2}>>m_{\phi}$. However, the dynamics is very similar to
  that in the potential (\ref{pot-1}), and so is the Q-ball
  formation. When we seek for the allowed region in the parameter
  space, the difference between $F^{1/2}$ and $m_{\phi}$ leads to the
  different conclusion. See Sect.~\ref{evap}.} 

Although we are considering the gauge mediation model, the flat
directions are also lifted by the gravity mediation mechanism, since
the gravity effects always exist. We can write in the form
\begin{equation}
    \label{pot-grav}
    V_{grav} = m_{grav}^2|\Phi|^2 
             = m_{3/2}^2 \left[ 1+K\log\left(\frac{|\Phi|^2}{M^2}
                             \right)\right]|\Phi|^2,
\end{equation}
where the $K$ term is the one-loop corrections \cite{EnMc1}, and
$M=2.4\times 10^{18}$ GeV is the reduced Planck mass. Since the
gravitino mass $m_{3/2}$ is much smaller than $\sim 1$ TeV, which is
suggested by the gravity-mediated SUSY breaking scenario, this term
will dominate over the gauge-mediation effects only at very large
scales \cite{DeMoMu,KK3}.

The flat directions may also be lifted by nonrenormalizable terms. 
When the superpotential is $W=\lambda\Phi^n/(n M^{n-3})$, the terms  
\begin{eqnarray}
    V_{NR} & = & \lambda^2\frac{|\Phi|^{2n-2}}{M^{2n-6}}, \\ \nonumber
    V_{A}  & = & \lambda A_{\lambda} \frac{\Phi^n}{M^{n-3}} + h.c.,
\end{eqnarray}
are added to the effective potential for the flat directions, where
$V_A$ denotes A-terms and we assume vanishing cosmological constant to
obtain them, so $|A_{\lambda}|\simeq m_{3/2}$. 

In addition to the terms above, there are those terms which depends on
the Hubble parameter $H$, during inflation and inflaton oscillation
stage which starts just after inflation. These read as
\begin{eqnarray}
    V_{H}  & = & -c_H H^2|\Phi|^2, \\
    \label{Aterm}
    V_{AH} & = & \lambda a_H H \frac{\Phi^n}{M^{n-3}} + h.c.,
\end{eqnarray}
where $c_H$ is a positive constant and $a_H$ is a complex constant with
a different phase from $A_{\lambda}$ in order for the AD mechanism to
work; we need a large initial amplitude of the field and a kicking
force for the phase rotation, which leads to the baryon number
creation. 

Now we are going to consider thermal effects on the effective
potential for the AD field. The flat directions can be lifted by
thermal effects of the SUSY breaking because of the difference of the
statistics between bosons and fermions. If the AD field couples
directly to particles in thermal bath, its potential receives thermal
mass corrections 
\begin{equation}
    V_T^{(m)} = c_T^{(1)} T^2 |\Phi|^2,
\end{equation}
where $c_T$ is proportional to the degrees of freedom of the thermal
particles coupling directly to the AD field. Therefore, if the
amplitude of the AD field is large enough, this term will vanish,
because the particles coupled to the AD field get very large effective 
mass, $m_{eff} \sim f|\Phi| \gg T$, where $f$ is a gauge or Yukawa
coupling constant between those particles and the AD field, so they
decouple from thermal bath. If we consider the formation of large Q
balls, the value of the AD field should be very large, and we can
neglect this term in the most situations. 

On the other hand, there is another thermal effect on the potential. 
This effect comes from indirect couplings of thermal particles to the 
AD field, mediated by heavy particles which directly couple to the 
AD field, and it reads as
\begin{equation}
    V_T^{(2)}= c_T^{(2)} T^4 \log \frac{|\Phi|^2}{T^2},
\end{equation}
where $c_T^{(2)}\sim \alpha^2(T)$ \cite{AnDi,FuHaYa}, where 
$\alpha(T)=g^2(T)/4\pi$, estimated at the temperature $T$. We will
combine these two terms in one form as 
\begin{equation}
    V_T = T^4 \log \left( 1+\frac{|\Phi|^2}{T^2}\right).
\end{equation}
Therefore, comparing with Eq.(\ref{pot-1}), we can regard the
effective mass as
\begin{equation}
    m(T) = \left\{
    \begin{array}{ll}
        m_{\phi} & (T<m_{\phi}) \\[2mm]
        T & (T>m_{\phi}) \\
    \end{array} \right.,
\end{equation}
and the effective potential as
\begin{equation}
    \label{pot-gen}
    V(\Phi) = m^4(T) \log \left( 1+\frac{|\Phi|^2}{m^2(T)}\right).
\end{equation}

As can be seen later, the Q-ball formation takes place very rapidly,
so the time dependence on $T$ is less effective, since 
$T \propto t^{-1/4}$ during inflaton-oscillation dominated universe.
Therefore, we can regard $T$ as constant and apply the numerical
results obtained for $m(T)=m_{\phi}$ to the case for $m(T)=T$ by
rescaling the variables to be dimensionless with respect to $T$.

\section{Affleck-Dine mechanism}
\label{AD-mech}
It was believed that the Affleck-Dine (AD) mechanism works as follows
\cite{AfDi,DiRaTh}. During inflation, the AD field are trapped at the
value determined by the following conditions: 
$V_H'(\Phi) \sim V_{NR}'(\Phi)$ and $V_{AH}'(\Phi) \sim 0$. Therefore, 
we get 
\begin{eqnarray}
    \label{AD-min}
    & & \phi_{min} \sim (HM^{n-3})^{1/(n-2)}, \nonumber \\
    & & \sin[n\theta_{min}+arg(a_H)] \sim 0,
\end{eqnarray}
where $c_H,\lambda \sim 1$ are assumed, and 
$\Phi=(\phi e^{i \theta})/\sqrt{2}$. After inflation, the inflaton
oscillates around one of the minimum of its effective potential, and
the energy of its oscillation dominates the universe. During that
time, the minimum of the potential of the AD field are adiabatically
changing until it rolls down rapidly when $H \simeq \omega\equiv
|V'(\phi)/\phi|^{1/2}$. Then the AD field rotates in its potential,
producing the baryon number $n_B = \dot{\theta} \phi^2$. After the AD
field decays into quarks, baryogenesis completes. 

In order to estimate the amount of the produced baryon number, let us
assume that the phases of $A_{\lambda}$ and $a_H$ differ of order
unity. Then the baryon number can be estimated as
\begin{eqnarray}
    n_B & \sim & H^{-1} \frac{\partial V_A}{\partial \phi}\phi 
          \sim H^{-1} V_A \nonumber \\
    & \sim &\omega^{-1} m_{3/2} \frac{\phi^n}{M^{n-3}}
      \sim   \left( \frac{m_{3/2}}{\omega} \right) \omega\phi^2
      = \varepsilon n_B^{(max)},
\end{eqnarray}
where $\varepsilon \equiv (m_{3/2}/\omega)$, 
$n_B^{(max)} \equiv \omega\phi^2$, and $H \sim \omega$ is used in the
second line. Notice that the contribution from the Hubble A-term is at 
most comparable to this. When $\varepsilon=1$, the trace of the motion
of the AD field in the potential is circular orbit. If $\varepsilon$
becomes smaller, the orbit becomes elliptic, and finally the field is
just oscillating along radial direction when $\varepsilon=0$. We call
$\varepsilon$ as the ellipticity parameter below.

When the logarithmic potential (\ref{pot-gen}) is dominant, 
$\omega \sim m^2/\phi$, so the ellipticity parameter is 
\begin{equation}
    \varepsilon \sim \frac{m_{3/2}\phi_0}{m^2(T)}.
\end{equation}
For $\varepsilon\sim 1$, it is necessary to have 
\begin{equation}
    m(T) \sim (m_{3/2}^{n-1}M^{n-3})^{1/(2n-4)},
\end{equation}
where we use Eq.(\ref{AD-min}) with $H\sim\omega\sim m^2(T)/\phi_0$.
When $m(T) \sim T$, the conditions for the AD field to rotate
circularly are
\begin{equation}
    \label{T-cir}
    T_{cir} \sim \left\{
      \begin{array}{ll}
          \ds{(m_{3/2}^3M)^{1/4} \sim 10^2 
               \left( \frac{m_{3/2}}{{\rm MeV}}\right)^{3/4} {\rm GeV}}
          & (n=4), \\[2mm]
          \ds{(m_{3/2}^5M^3)^{1/8} \sim 10^5 
               \left( \frac{m_{3/2}}{{\rm MeV}}\right)^{5/8} {\rm GeV}}
          & (n=6).
      \end{array}
      \right.
\end{equation}

On the other hand, when the potential is dominated by $V_{grav}$,
$\omega\simeq m_{3/2}$. Then $\varepsilon \simeq 1$ always holds.
Of course, this conclusion does not necessarily true, if the A-terms
do not come from the nonrenormalization term in the superpotential,
such as in Eq.(\ref{Aterm}). $\varepsilon$ is generally arbitrary, if
the origin of the A-terms is in the K\"{a}hler potential, which we
have less knowledge.

\section{Instabilities of Affleck-Dine field}
In the Affleck-Dine mechanism, it is usually assumed that the field
rotates in its potential homogeneously to produce the baryon
number. However, the field feels spatial instabilities, and they grow
to nonlinear to become clumpy objects, Q balls \cite{KuSh}. In this
section, we study the instability of the field dynamics both
analytically and numerically. 

For the discussion to be simple, we consider the field dynamics in the 
logarithmic potential only when the gauge-mediated SUSY breaking
effects dominate. This is enough to investigate the whole
dynamics, since fluctuations are developed when the field stays in
this region of the effective potential. We write a complex field as 
$\Phi = (\phi e^{i \theta})/\sqrt{2}$, and decompose into homogeneous 
part and fluctuations: $\phi \rightarrow \phi +\delta\phi$ and
$\theta \rightarrow \theta + \delta\theta$. Then the equations of
motion of the AD field read as \cite{KuSh,EnMc2}
\begin{eqnarray}
    \ddot{\phi} + 3H\dot{\phi} - \dot{\theta}^2\phi
      + \frac{m^2\phi}{1+\frac{\phi^2}{2m^2}} & = & 0, \\
    \label{theta-eq}
    \phi\ddot{\theta} + 3H\phi\dot{\theta} 
      + 2\dot{\phi}\dot{\theta} & = & 0,
\end{eqnarray}
for the homogeneous mode, and
\begin{eqnarray}
    \label{eom-fl}
    \delta\ddot{\phi} + 3H\delta\dot{\phi}
     - 2\dot{\theta}\phi\delta\dot{\theta} - \dot{\theta}^2\delta\phi
     -\frac{\nabla^2}{a^2}\delta\phi + V''(\phi)\delta\phi & = & 0, 
     \nonumber \\
     \phi\delta\ddot{\theta} 
       + 3H\phi\delta\dot{\theta}
       + 2(\dot{\phi}\delta\dot{\theta} 
           +\dot{\theta}\delta\dot{\theta})
       -2\frac{\dot{\phi}}{\phi}\dot{\theta}\delta\phi
       -\phi\frac{\nabla^2}{a^2}\delta\theta & = & 0,
\end{eqnarray}
for the fluctuations, where $a$ is the scale factor of the universe,
and
\begin{equation}
    V''(\phi) = m^2 \frac{1-\frac{\phi^2}{2m^2}}
                   {\left(1+\frac{\phi^2}{2m^2}\right)^2}
   \simeq - \frac{2m^4}{\phi^2},
\end{equation}
where we assume $\phi \gg m$ in the last step. Equation
(\ref{theta-eq}) represents the conservation of the charge (or number) 
within the physical volume: $\dot{\theta}\phi^2a^3=const.$

Now let us forget about the cosmic expansion, since it only makes our 
analysis be complicated. Thus, $a=const.$ here. It is natural to take
a circular orbit of the field motion in the potential. (For a 
noncircular orbit, it is difficult to discuss analytically, so we will
show numerical results later.) Then we can take $\phi=\phi_0=const.$,
and the phase velocity  
$\dot{\theta} = (V'/\phi)^{1/2} \simeq \sqrt{2}m^2/\phi_0$. We
seek for the solutions in the form
\begin{equation}
    \delta\phi = \delta\phi_0 e^{\alpha t+ikx}, \qquad
    \delta\theta = \delta\theta_0 e^{\alpha t+ikx}.
\end{equation}
If $\alpha$ is real and positive, these fluctuations grow
exponentially, and go nonlinear to form Q balls. Inserting these forms 
into Eqs.(\ref{eom-fl}), we get the following condition for nontrivial 
$\delta\phi_0$ and $\delta\theta_0$,
\begin{equation}
   \label{det}
   \left|
      \begin{array}{ccc}
          \ds{\alpha^2+\frac{k^2}{a^2}-\frac{4m^4}{\phi_0^2}}
          & & \ds{-2\sqrt{2}m^2\alpha} \\[4mm]
          \ds{\frac{2\sqrt{2}m^2}{\phi_0^2}\alpha}
          & & \ds{\alpha^2+\frac{k^2}{a^2}}
    \end{array}
    \right| = 0.
\end{equation}
This equation can be simplified to be
\begin{equation}
    \alpha^4 
     + 2\left( \frac{k^2}{a^2} + \frac{2m^4}{\phi_0^2} \right)\alpha^2 
     + \left( \frac{k^2}{a^2} - \frac{4m^4}{\phi_0^2} \right) k^2 = 0.
\end{equation}
In order for $\alpha$ to be real and positive, we must have
\begin{equation}
    \left( \frac{k^2}{a^2} - \frac{4m^4}{\phi_0^2} \right) 
           \frac{k^2}{a^2} < 0.
\end{equation}
Therefore, we obtain the instability band for the fluctuations as
\begin{equation}
    0 < \frac{k}{a} < \frac{2m^2}{\phi_0}.
\end{equation}
We can easily derive that the most amplified mode appears at
$(k/a)_{max} = (3/2)^{1/2}m^2/\phi_0$, and the largest growth
factor is 
$\alpha_{max} \equiv \alpha(k_{max}) = m^2/(\sqrt{2}\phi_0)$.

It is easier to decompose a complex field into its real and imaginary
parts for analyzing the dynamics of the field when its motion is
noncircular. For the numerical calculation, it is convenient to take
all the variables to be dimensionless, so we normalize as
$\varphi=\phi/m$, $\tilde{k}=k/m$, $\tau = mt$, and $\xi = mx$.
Writing $\varphi = (\varphi_1 + i \varphi_2)/\sqrt{2}$, we get the
equations for the homogeneous mode as
\begin{eqnarray}
    \varphi_1''+\frac{\varphi_1}{1+\frac{\varphi_1^2+\varphi_2^2}{2}}
    & = & 0, \nonumber \\
    \varphi_2''+\frac{\varphi_2}{1+\frac{\varphi_1^2+\varphi_2^2}{2}}
    & = & 0, 
\end{eqnarray}
and, for the fluctuations,
\begin{equation}
    \left[\frac{d^2}{d\tau^2} + \frac{\tilde{k}^2}{a^2} 
      + V_{ij}\right]
    \left(\begin{array}{c}
            \ds{\delta\varphi_1} \\
            \ds{\delta\varphi_2}
          \end{array} \right) = 0,
\end{equation}
where $V_{ij}$ denotes the second derivative with respect to
$\varphi_i$ and $\varphi_j$, and explicitly written as
\begin{eqnarray}
    & &
    V_{11} = \frac{1-\frac{\varphi_1^2-\varphi_2^2}{2}}
            {\left(1+\frac{\varphi_1^2+\varphi_2^2}{2}\right)^2}, 
    V_{22} = \frac{1+\frac{\varphi_1^2-\varphi_2^2}{2}}
            {\left(1+\frac{\varphi_1^2+\varphi_2^2}{2}\right)^2}, 
    \nonumber \\
    & &
    V_{12} = V_{21} = \frac{\varphi_1\varphi_2}
            {\left(1+\frac{\varphi_1^2+\varphi_2^2}{2}\right)^2}. 
\end{eqnarray}

\begin{figure}[t!]
\centering
\hspace*{-7mm}
\leavevmode\epsfysize=6cm \epsfbox{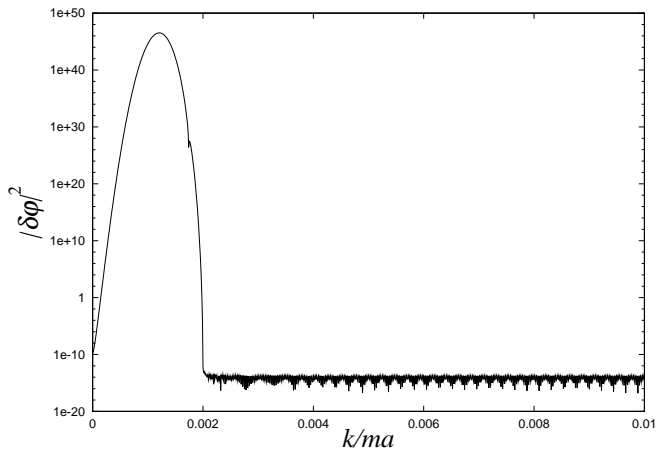}\\[2mm]
\caption[fig1]{\label{fig1} 
Instability band for circular orbit $\varepsilon=1$ in the logarithmic 
potential (\ref{pot-1}).}
\end{figure}

\begin{figure}[t!]
\centering
\hspace*{-7mm}
\leavevmode\epsfysize=6cm \epsfbox{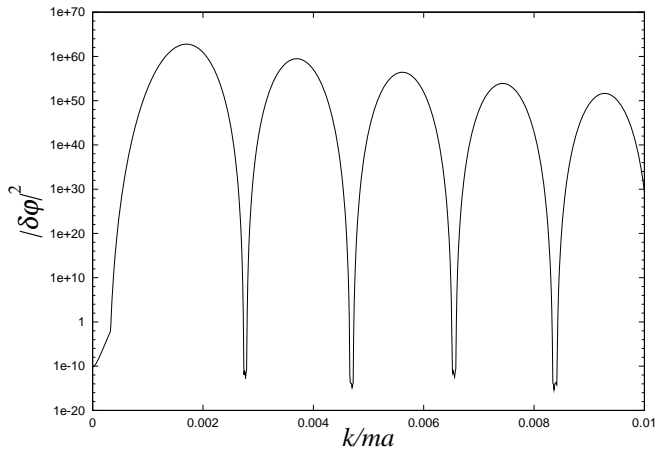}\\[2mm]
\caption[fig2]{\label{fig2} 
Instability bands for radial oscillation $\varepsilon=0$ in the
logarithmic potential (\ref{pot-1}).}
\end{figure}

We show two typical situations for the field
evolution. Figure~\ref{fig1} shows the result for the initial
conditions, $\varphi_1(0)=10^3$ and $\varphi_2'(0)=\sqrt{2}$, where
the prime denotes the derivative with respect to $\tau$. This
corresponds to the circular orbit for the motion of the homogeneous
mode. We can see that the instability band coincides exactly with that
obtained analytically, since the upper bound of the instability band
is $2/\varphi(0)=0.002$ in dimensionless units. On the other hand,
Figure~\ref{fig2} shows the results for $\varphi_1(0)=10^3$ and
$\varphi_2'(0)=0$, which corresponds to the case that the homogeneous
field is just oscillating along the radial direction. Almost all the
modes are in the instability bands, but they have quasi-periodical
structures. This shows some features of the parametric
resonance. Anyway, the most important consequence is the position of
the most amplified mode, since it corresponds to the typical size of
produced Q balls. Comparing with these two cases, a larger Q ball
should form in the former (circular orbit) case, and the size will be
about twice as large as that of the latter case, because the most
amplified mode $k_{max}$ for the circular orbit case is about twice
smaller than that for the just-oscillation case.

We also see situations between circular orbit ($\varepsilon=1$) and
just-oscillation ($\varepsilon=0$), where $\varepsilon$ represents the 
ellipticity of the orbit and defined by
$\varphi_2'(0)=\sqrt{2}\varepsilon$. Higher instability bands become
narrower and disappear as $\varepsilon \rightarrow 1$, and the lowest 
band also becomes narrower.

Now let us move on to the fluctuations grown in the potential
(\ref{pot-grav}) where the gravity mediation effects dominate. 
Instabilities grow if $K$ is negative, which is realized when the
quantum corrections to the potential is dominated by the gaugino
loops. We analyze this situation in Ref.~\cite{KK2}, so we only show
the results here. The instability modes are in the range
$0<(k/a)^2<3|K|m_{3/2}^2$. The most amplified mode is $(k_{max}/a)^2 =
3|K|m_{3/2}^2/2$, and the maximum growth rate is 
$\alpha_{max} = 3|K|m_{3/2}/4$ \cite{KK2}.

\section{Q-ball formation}
Generally, Q balls are produced during the rotation of the AD
field. In Ref.~\cite{KK1}, we investigated the Q-ball formation in
the gauge-mediation scenario on the three dimensional lattices, and
found that it actually occurs. In there, we used the effective
potential of the AD field as
\begin{equation}
    \label{pot-2}
    V(\Phi)=m_{\phi}^4\log\left(1+\frac{|\Phi|^2}{m_{\phi}^2}\right) 
      + \frac{\lambda^2}{M^2}|\Phi|^6,
\end{equation}
and took the initial conditions for the homogeneous mode as
$\tau(0)=100$, $\varphi_1(0)=A$, $\varphi_1'(0)=0$, $\varphi_2(0)=0$,
and $\varphi_2'(0)=B$, where we varied $A$ and $B$ in some ranges, and 
added small random values representing fluctuations. However,
at the initial amplitude which we took, the potential is dominated by
the nonrenormalizable term, and for our choice of the initial time,
the cosmic expansion is weaker than the realistic case. Thus, we could
only find some restricted relations between initial values and charges
of Q balls produced.

Therefore, we use only the first term of Eq.(\ref{pot-2}), or
equivalently, Eq.(\ref{pot-gen}) with $m(T)=const.$ for the
potential, and take initial conditions as
\begin{equation}
    \begin{array}{ll}
        \ds{\varphi_1(0) = \varphi_0 +\delta_1,} & 
        \ds{\varphi_1'(0) = \delta_2,} \\
        \ds{\varphi_2(0) = \delta_3,} &
        \ds{\varphi_2'(0) = \varepsilon\sqrt{2} + \delta_4,} \\[3mm]
        \ds{\tau(0)=\frac{2}{3h}=\frac{\sqrt{2}}{3}\varphi_0,} &
    \end{array}
\end{equation}
where all variables are rescaled to be dimensionless parameters,
$h=H/m$, and $\varepsilon$ represents how circular the orbit of 
the field motion is $\varepsilon=1$ for the circular orbit and 
$\varepsilon=0$ for the radial oscillation. $\delta$'s are
fluctuations which originate from the quantum fluctuations of the AD
field during inflation, and their amplitudes are estimated as 
$10^{-7}-10^{-10}$ compared with corresponding homogeneous modes. The
initial time is taken to be at the time when $H(t)=\omega$, which is
the condition that the AD field starts its rotation.

\begin{figure}[t!]
\centering
\hspace*{-7mm}
\leavevmode\epsfysize=6cm \epsfbox{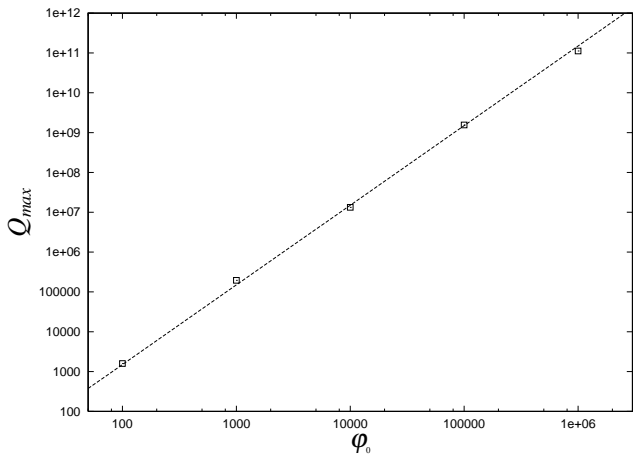}\\[2mm]
\caption[fig3]{\label{fig3} 
$Q-\varphi_0$ relation in the logarithmic potential on one-dimensional
lattices. Dashed line denotes $Q_{max}=0.15 \varphi_0^2$.}
\end{figure}

\begin{figure}[t!]
\centering
\hspace*{-7mm}
\leavevmode\epsfysize=6cm \epsfbox{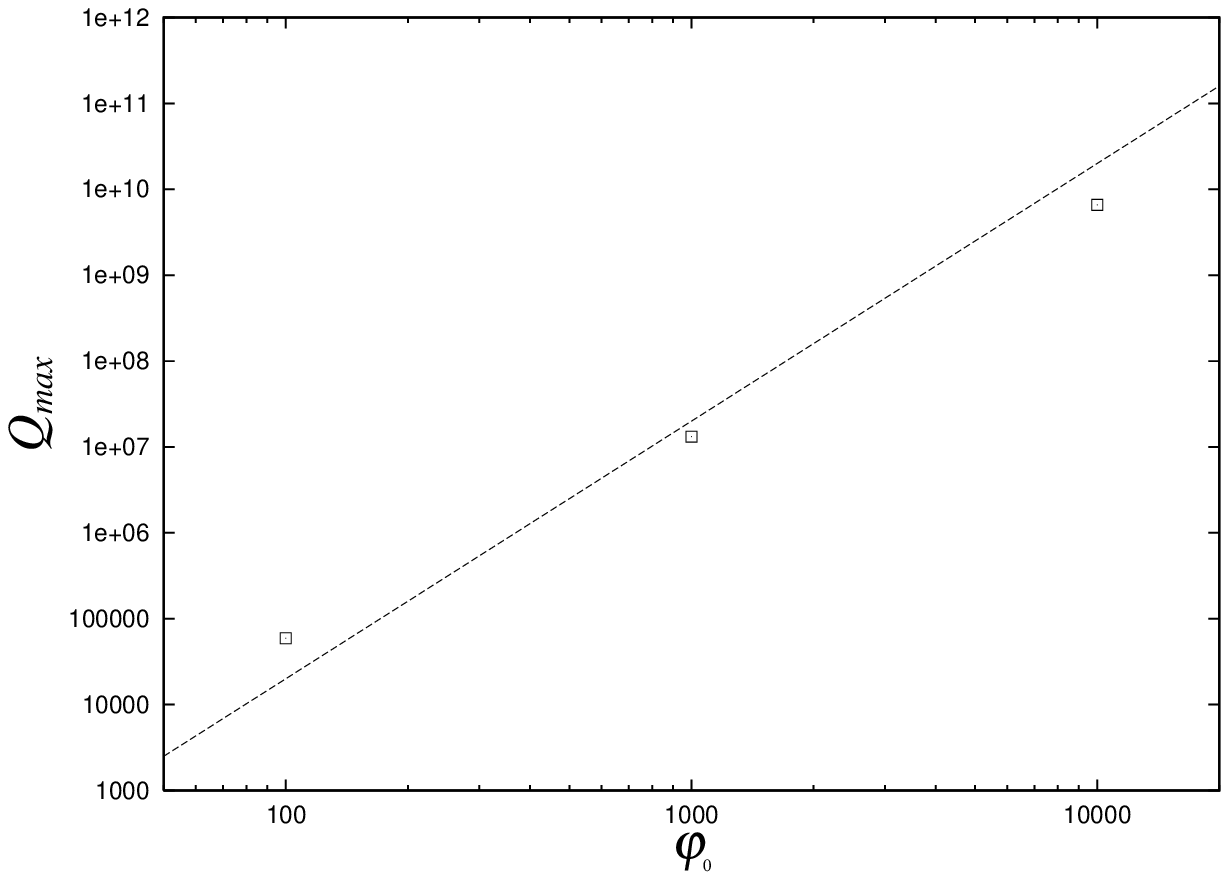}\\[2mm]
\caption[fig4]{\label{fig4} 
$Q-\varphi_0$ relation in the logarithmic potential on two-dimensional
lattices. Dashed line denotes $Q_{max}=2\times10^{-2}\varphi_0^3$.}
\end{figure}

\begin{figure}[t!]
\centering
\hspace*{-7mm}
\leavevmode\epsfysize=6cm \epsfbox{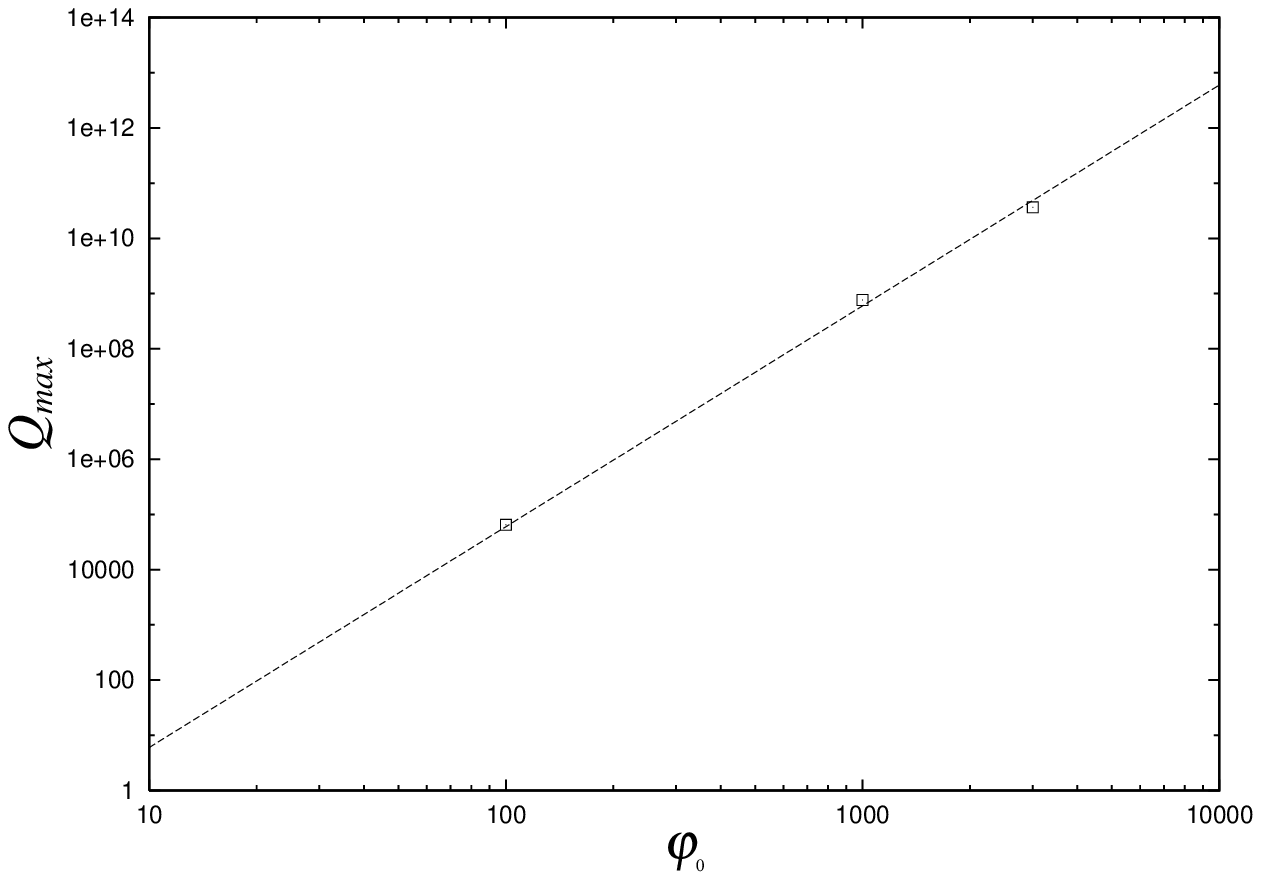}\\[2mm]
\caption[fig5]{\label{fig5} 
$Q-\varphi_0$ relation in the logarithmic potential on 
three-dimensional lattices. Dashed line denotes
$Q_{max}=6\times10^{-4}\varphi_0^4$.} 
\end{figure}

We calculate in one, two, and three dimensional lattices. In the 
first place, we set $\varepsilon=1$, and vary the initial amplitude of 
the AD field $\varphi_0$. It is usually considered that $\varepsilon$
may be very small in the gauge mediation scenario, because the
gravitino mass $m_{3/2}$ is very small so that the amplitude of
A-terms may be small. However, as we have shown above, 
$\varepsilon=1$ case can be achieved in certain situations. Figures
\ref{fig3} - \ref{fig5} show the dependence of the maximum Q-ball
charge produced upon the initial amplitude $\varphi_0$ for one, two,
and three dimensional lattices, respectively.

We find the relation between the charge of the Q ball and the initial 
amplitude of the AD field:
\begin{equation}
    \label{charge-phi}
    Q_{max} = \beta_d \varphi_0^{1+d}, \qquad d=1,2,3,
\end{equation}
where $\beta_d$'s are some numerical factors; $\beta_1\approx 0.1$,
$\beta_2 \approx 0.02$, and $\beta_3 \approx 6\times 10^{-4}$.
This relation can be understood as follows. Since the wavelength of
the most amplified mode is as large as the horizon size, we can regard
that one Q ball is created in the horizon size at a creation time. If
we assume that Q balls are created just after the AD field starts its
rotation, the horizon size is 
$H^{-1} \sim \omega^{-1} \sim \phi_0/m^2$. Therefore, the charge of a
Q ball is
\begin{equation}
    Q \sim H^{-d} n_{\phi} \sim \omega^{-d} \omega\phi_0^2
      \sim m^{3-d}\left(\frac{\phi_0}{m}\right)^{1+d}.
\end{equation}
This corresponds to the results of numerical calculations: 
$Q \propto \varphi_0^{1+d}$. The prefactors $\beta_d$ cannot be
determined by the above analytical estimation. Actually, the formation
time is a little bit later than the time when the rotation of the AD
field starts, and the number of Q balls in the horizon size is more
than one. Therefore, we will use $\beta_3 \approx 6\times10^{-4}$ for
the estimation of Q-ball charge when we need later.

\begin{figure}[t!]
\centering
\hspace*{-7mm}
\leavevmode\epsfysize=6cm \epsfbox{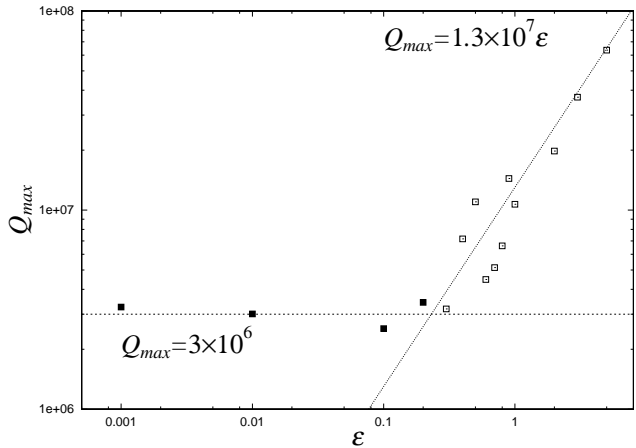}\\[2mm]
\caption[fig6]{\label{fig6} 
$\varepsilon$-dependence of Q-ball charge in the logarithmic potential 
on one-dimensional lattices. The dotted line denotes
$Q_{max}\simeq 1.3\times10^7 \varepsilon$. On the other hand, the
dashed line denotes the constant line $Q \simeq 3\times 10^6$. Closed
squares denote the production of both positive and negative Q balls
with charges of opposite signs and the same order of magnitude.}
\end{figure}

Now let us see how the Q-ball formation occurs for 
$\varepsilon \rightarrow 0$. The charge of produced Q balls decreases
with linear dependence on the charge density of the AD field, as shown 
in Ref.~\cite{KK1}, thus proportional to $\varepsilon$. When the
angular velocity of the AD field becomes small enough, both Q balls
with positive and negative charges are produced \cite{KK1,KK2,KK4}.
Therefore, the charge of the Q ball will be independent of small
enough $\varepsilon$. Using numerical calculation, we find that the
dependence of the Q-ball charge on $\varepsilon$ shows exactly the
same features as expected. In Fig.~\ref{fig6}, we plot the Q-ball
charge on one dimensional lattices. We can see that $Q_{max}$ is
constant for small $\varepsilon$ where both positive and negative Q
balls with the charges of the same order of magnitude with opposite
signs are produced, while linearly dependent on $\varepsilon$ around
$\varepsilon \sim 1$, where only positive Q balls are formed (smaller
negative Q balls can be produced, but their charges are an order of
magnitude or more smaller than the largest positive Q balls). We 
thus obtain the formula for the charge of the produced Q balls as
\begin{equation}
    Q_{1D} \simeq \left\{
      \begin{array}{ll}
          \ds{3\times10^6,} & \quad \ds{\varepsilon\lesssim0.2}, \\
          \ds{1.3\times10^7 \varepsilon,} & \quad 
              \ds{\varepsilon\gtrsim0.2},
      \end{array}
      \right.
\end{equation}
in one dimension.

\begin{figure}[t!]
\centering
\hspace*{-7mm}
\leavevmode\epsfysize=6cm \epsfbox{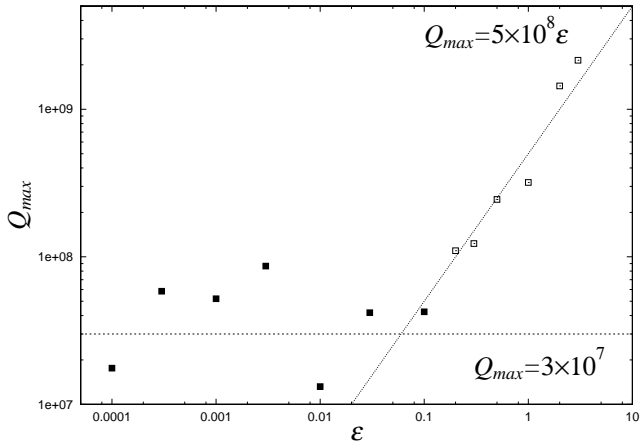}\\[2mm]
\caption[fig7]{\label{fig7} 
$\varepsilon$-dependence of Q-ball charge in the logarithmic potential 
on three-dimensional lattices. The dotted line denotes
$Q_{max}\simeq 5\times10^8 \varepsilon$. On the other hand, the
dashed line denotes the constant line $Q \simeq 3\times 10^7$.Closed
squares denote the production of both positive and negative Q balls
with charges of opposite signs and the same order of magnitude.}
\end{figure}

Similarly, we plot the Q-ball charge on three dimensional lattices in
Fig.~\ref{fig7}. Also we can see that $Q_{max}$ is constant for small
$\varepsilon$ where both positive and negative Q balls are produced,
while linearly dependent on $\varepsilon$ around $\varepsilon \sim 1$,
where only positive Q balls are formed. Therefore, we obtain the
following formula for the charge of the produced Q balls:
\begin{equation}
    Q_{3D} \simeq \left\{
      \begin{array}{ll}
          \ds{3\times10^7,} & \quad \ds{\varepsilon\lesssim0.06}, \\
          \ds{5\times10^8 \varepsilon,} & \quad 
             \ds{\varepsilon\gtrsim0.06},
      \end{array}
      \right.
\end{equation}

The difference between the charges for $\varepsilon \sim 1$ and 
$\varepsilon \ll 1$ can be explained by the size of the produced Q
balls. As seen in the previous section, the most amplified mode
$k_{max}$ of the fluctuations for the circular orbit ($\varepsilon=1$)
is about twice as small as that of just-oscillation along the radial
direction ($\varepsilon=0$). Therefore, twice larger Q balls are
produced for $\varepsilon=1$ in one dimension, while about an order
bigger ones are formed in three dimension, which exactly correspond to 
the difference between charges for $\varepsilon \sim 1$ and
$\varepsilon \ll 1$.

At very large amplitudes of the AD field, the gravity mediation
effects for SUSY breaking will dominate, and the `new' type of stable
Q balls are produced \cite{KK3}. As we see above, the potential is
dominated by the form 
\begin{equation}
    \label{grav-pot}
    V(\Phi) = m_{3/2}^2\left[1+K\log
        \left(\frac{|\Phi|^2}{M^2}\right)\right]|\Phi|^2.
\end{equation}
In this case, the curvature of the potential does not depend on the
amplitude so much, so the AD field starts its rotation when 
$H \simeq m_{3/2}$. Thus, the initial time can be taken as 
$\tau(0) = 2/3$ if we rescale variables with respect to $m_{3/2}$
when we make them dimensionless. 

\begin{figure}[t!]
\centering
\hspace*{-7mm}
\leavevmode\epsfysize=6cm \epsfbox{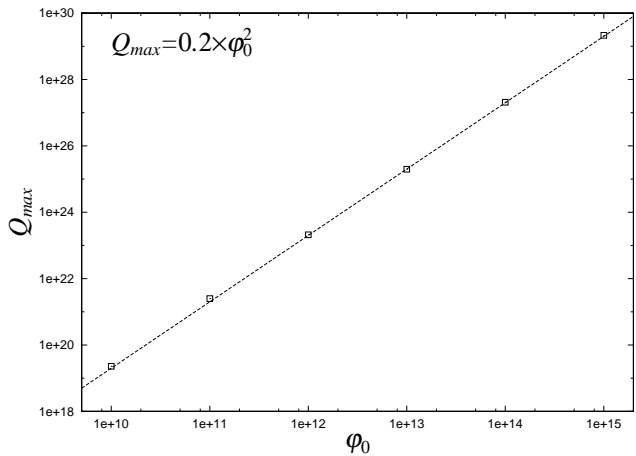}\\[2mm]
\caption[fig8]{\label{fig8} 
$Q-\varphi_0$ relation in the potential (\ref{grav-pot}) \break on 
one-dimensional lattices. Dashed line denotes \break
$Q_{max}=0.2\times\varphi_0^2$.}
\end{figure}

\begin{figure}[t!]
\centering
\hspace*{-7mm}
\leavevmode\epsfysize=6cm \epsfbox{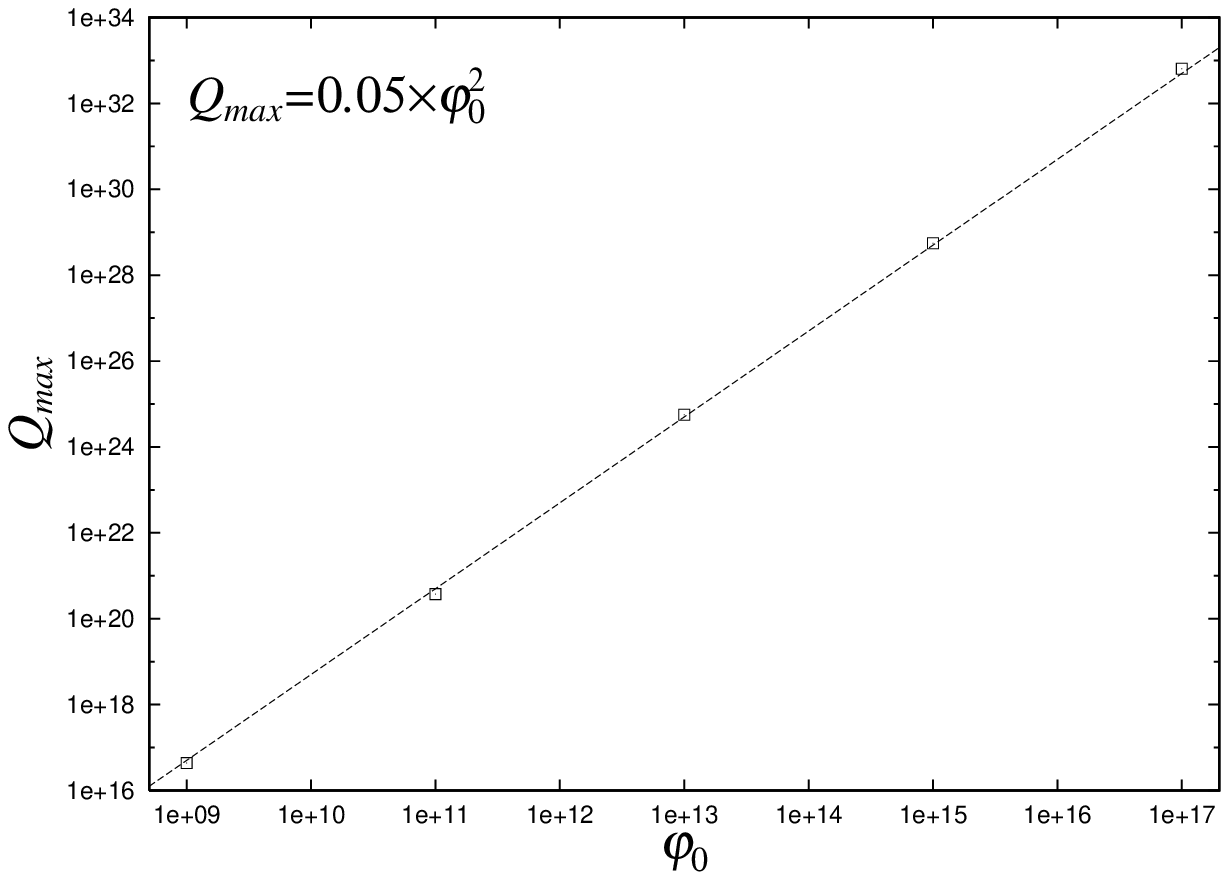}\\[2mm]
\caption[fig9]{\label{fig9} 
$Q-\varphi_0$ relation in the potential (\ref{grav-pot}) \break on 
two-dimensional lattices. Dashed line denotes \break
$Q_{max}=0.05\times\varphi_0^2$.}
\end{figure}

\begin{figure}[t!]
\centering
\hspace*{-7mm}
\leavevmode\epsfysize=6cm \epsfbox{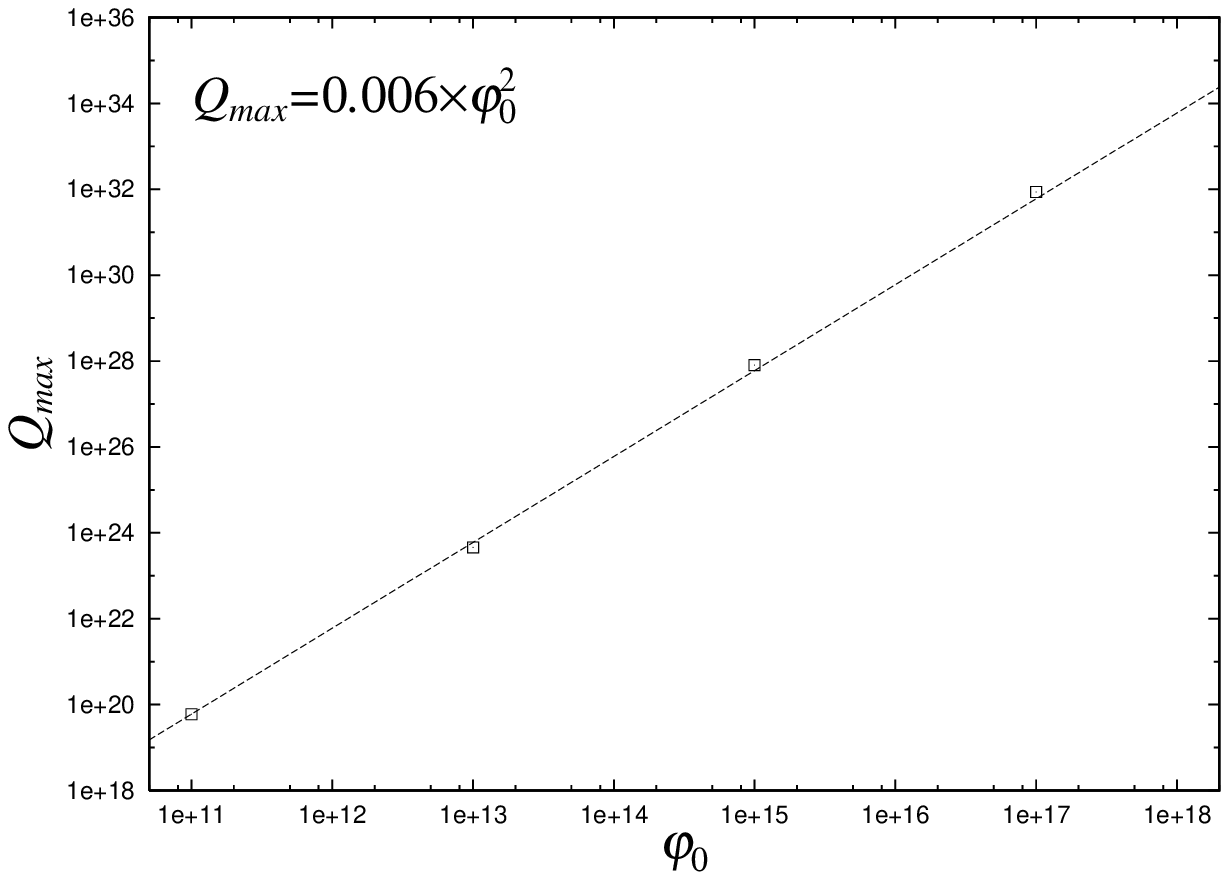}\\[2mm]
\caption[fig10]{\label{fig10} 
$Q-\varphi_0$ relation in the potential (\ref{grav-pot}) on 
three-dimensional lattices. Dashed line denotes 
$Q_{max}=0.006\times\varphi_0^2$.}
\end{figure}

We have simulated the dynamics of the AD field on one, two, and three
dimensional lattices, and find the formation of Q balls. Here, we take 
$K=-0.01$. (Actually, its value is estimated in the range from $-0.01$ 
to $-0.1$ \cite{EnMc1,EnJoMc}.) We plot the largest Q-ball charge in
the function of the initial amplitude $\varphi_0$ for one, two,
and three dimensional lattices in Figs.~\ref{fig8}, \ref{fig9}, and
\ref{fig10}, respectively.

Similar to the logarithmic potential cases, we find the relation
between the charge of the Q ball and the initial amplitude of the AD
field as 
\begin{equation}
    Q_{max} = \tilde{\beta}_d \varphi_0^{2}, \qquad d=1,2,3,
\end{equation}
where $\tilde{\beta}_d$'s are some numerical factors; 
$\tilde{\beta}_1\approx 0.2$, $\tilde{\beta}_2 \approx 0.05$, and  
$\tilde{\beta}_3 \approx 0.006$. This relation can be understood by
the similar argument as the logarithmic potential case. The
difference is that the time when the AD field starts oscillation and
the most amplified mode (or instability band) enters the
horizon do not coincide. Therefore, the charge of a Q ball is
\cite{KK2}
\begin{eqnarray}
    Q \sim H_f^{-d} m_{3/2}\phi_f^2
      & \sim & (|K|^{1/2}m_{3/2})^{-d} m_{3/2}|K|\phi_0^2 \nonumber \\
      & \sim & |K|^{1-d/2}m_{3/2}^{3-d}
            \left(\frac{\phi_0}{m_{3/2}}\right)^2,
\end{eqnarray}
where the subscript `$f$' denotes the values at the Q-ball formation
time. This corresponds to the results of numerical calculations: 
$Q \propto \varphi_0^{2}$. The prefactors $\tilde{\beta}_d$ cannot be
determined by the above analytical estimation. Actually, the formation
time is a little bit later than the time when the mode which will be
mostly amplified enters the horizon, and the number of the Q balls in
the horizon size is more than one. Therefore, we use $\tilde{\beta}_3
\approx 6\times10^{-3}$ for the estimation of the Q-ball charge when
we need later.

\section{Charge evaporation from Q balls in high temperature}
\subsection{Usual gauge-mediation type Q balls}
Since the Q-ball formation takes place nonadiabatically, (almost) all
the charges are absorbed into produced Q balls \cite{KK1,KK2}. Thus,
the baryon number of the universe cannot be explained by the remaining
charges outside Q balls in the form of the relic AD field. However, at
finite temperature, the minimum of the free energy of the AD field is
achieved for the situation that all the charges exist in the form of
free particles in thermal plasma \cite{LaSh}. This situation is
realized through charge evaporation from Q-ball surface
\cite{LaSh,EnMc2}. In spite of the fact that the complete evaporation
is the minimum of the free energy, the actual universe is filled with
the mixture of Q balls and surrounding free particles, since the
evaporation rate becomes smaller than the cosmic expansion rate at low 
temperatures. (As we mentioned in the Introduction, the free energy is 
minimized at the situation that some charges exist in the form of free
particles (in thermal plasma) and the rest stays inside Q balls, if
the Q-ball charge are large enough for the chemical equilibrium
\cite{LaSh}. In this case, the charge of the Q ball should be 
$Q \sim \eta_B^{-4} \sim 10^{40}$.) The nonadiabatic creation of Q
balls and the later charge evaporation takes place, since the time
scale of evaporation is much longer than that of the Q-ball 
formation. Notice that the charge of the Q ball is conserved for the
situation that the evaporation of charges is not effective. However,
the energy of the Q ball decreases as the temperature of the universe
decreases. This happens because the Q ball and surrounding plasma are
in thermal equilibrium: i.e., the Q balls and thermal plasma have the
same temperatures. Thus, the destruction process of the Q ball by the
collision of the thermal particles, called the dissociation
\cite{EnMc2} cannot occur in actual situations.

The rate of charge transfer from inside the Q ball to its outside is
determined by the diffusion rate at high temperature and the
evaporation rate at low temperature \cite{KK3}. When the
difference between the chemical potentials of the plasma and the Q
ball is small, chemical equilibrium is achieved and charges inside the
Q ball cannot come out \cite{BaJe}. Therefore, the charges in the
`atmosphere' of the Q ball should be taken away in order for further
charge evaporation. This process is determined by the diffusion. The
diffusion rate is given by \cite{BaJe}
\begin{equation}
    \label{diff-rate}
    \Gamma_{diff} \equiv \frac{dQ}{dt} \sim -4\pi D R_Q \mu_Q T^2
                                       \sim -4\pi A T, 
\end{equation}
where $D=A/T$ is the diffusion coefficient, and $A=4-6$, $\mu_Q \sim 
\omega$ is the chemical potential of the Q ball. On the other hand,
the evaporation rate is \cite{LaSh}
\begin{eqnarray}
    \label{evap-rate}
    \Gamma_{evap} \equiv \frac{dQ}{dt} 
    & \sim & -\zeta(\mu_Q-\mu_{plasma})T^2 4\pi R_Q^2 \nonumber \\
    & \sim & -4\pi\zeta \frac{T^2}{m(T)}Q^{1/4},
\end{eqnarray}
where $\mu_{plasma}\ll \mu_Q$ is the chemical potential of thermal
plasma, and $\mu_Q \sim \omega \sim m(T) Q^{-1/4}$ is used in the
second line. $m(T)$ and $\zeta$ change at $T=m_{\phi}$ as
\begin{equation}
   m(T)=\left\{
      \begin{array}{l}
          \ds{m_{\phi}} \\[3mm]
          \ds{T}
      \end{array}
    \right.,
    \quad
    \zeta=\left\{
      \begin{array}{cl}
          \ds{\left(\frac{T}{m_{\phi}}\right)^2} 
                    & \ds{(T<m_{\phi}),} \\[2mm]
          \ds{1}  & \ds{(T>m_{\phi}).}
      \end{array}
      \right.
\end{equation}
Therefore, we get 
\begin{equation}
    \label{evap-rate-2}
    \Gamma_{evap} = \frac{dQ}{dt} = \left\{
      \begin{array}{ll}
          \ds{-4\pi T Q^{1/4}} & \ds{(T>m_{\phi}),} \\
          \ds{-4\pi \frac{T^4}{m_{\phi}^3}Q^{1/4}} & 
                      \ds{(T<m_{\phi}).} 
      \end{array}
      \right.
\end{equation}

In order to see which rate is the bottle-neck for the process, let us
take the ratio $R\equiv \Gamma_{diff}/\Gamma_{evap}$. For
$T>m_{\phi}$, the ratio becomes $R = A Q^{-1/4}$. If $R<1$, the
diffusion rate is the bottle-neck for the charge transfer, and this
condition meets when the Q-ball charge is large enough as
\begin{equation}
    Q > 10^2 \left(\frac{A}{4}\right)^2.
\end{equation}
Since, as we will see shortly, we are interested in the dark matter Q
ball, the charge should be as large as $10^{24}$. More conservatively 
speaking, $Q\gtrsim 10^{12}$ for stable against the decay into
nucleons \cite{KuSh}. 

On the other hand, when $T<m_{\phi}$, the condition $R<1$ corresponds
to
\begin{equation}
    \label{crit-T}
    T > T_* \equiv A^{1/3} m_{\phi} Q^{-1/12},
\end{equation}
and the transition temperature $T_*$ is lower than $m_{\phi}$ for
large enough Q-ball charge, which is again the interesting range for
the dark matter Q balls.

We must know the time dependence of the temperature for estimating the 
total evaporated charge from the Q ball, which will be obtained by 
integrating Eqs.(\ref{diff-rate}) and (\ref{evap-rate}) [or
(\ref{evap-rate-2})]. Although thermal plasma (radiation) dominates
the energy density of the universe only after reheating, it exists
earlier, and this subdominant component will cause both the change of
the shape of the effective potential for the AD field and the charge
evaporation from Q balls. The temperature of the radiation before
reheating is 
\begin{equation}
    \label{T-br}
    T \sim (M^2\Gamma_IH)^{1/4} \sim (M T_{RH}^2 H)^{1/4},
\end{equation}
where $\Gamma_I$ is the decay rate of the inflaton field, and the
relation between the decay rate and reheating temperature, $T_{RH} 
\sim (M\Gamma_I)^{1/2}$, is used in the last equality. Of course, when 
reheating occurs at $H \sim \Gamma_I$, the temperature becomes as
large as the reheating temperature: $T \sim T_{RH}$. After reheating,
the universe is dominated by radiation, and the temperature evolves as 
$T \propto H^{1/2}$.

Since $t \propto T^{-4}$ before reheating, and $t\propto T^{-2}$ after
reheating, the diffusion rate with respect to temperature $T$ can be
directly obtained from Eq.(\ref{diff-rate}) as
\begin{equation}
    \left(\frac{dQ}{dT} \right)_{diff} \sim \left\{
      \begin{array}{ll}
          \ds{10 \frac{MT_{RH}^2}{T^4}} & \ds{(T>T_{RH}),} \\[2mm]
          \ds{10 \frac{M}{T^2}} & \ds{(T<T_{RH}).}
      \end{array}
      \right.
\end{equation}
On the other hand, the evaporation rate with respect to $T$ is a
little more complicated, and we can divide into four cases, depending
on the temperature compared with the reheating temperature and the
mass of the AD particle. The time dependence of $T$ changes at 
$T=T_{RH}$, and the rate $dQ/dt$ changes at $T=m_{\phi}$. Combining all
the effects, we get
\begin{equation}
     \left(\frac{dQ}{dT} \right)_{evap} \sim \left\{
      \begin{array}{ll}
          \ds{10 \frac{MT_{RH}^2}{T^4}Q^{1/4}} 
                       & \ds{(T>T_{RH},m_{\phi}),} \\[2mm]
          \ds{10 \frac{MT_{RH}^2}{m_{\phi}^3T}Q^{1/4}} 
                       & \ds{(T_{RH}<T<m_{\phi}),} \\[2mm]
          \ds{10 \frac{M}{T^2}Q^{1/4}} 
                       & \ds{(m_{\phi}<T<T_{RH}),} \\[2mm]
          \ds{10 \frac{MT}{m_{\phi}^3}Q^{1/4}} 
                       & \ds{(T<m_{\phi},T_{RH}).}
      \end{array}
      \right.
\end{equation}

We can now estimate the total evaporated charge from the Q ball.
According to the relation among $T_*$, $T_{RH}$, and $m_{\phi}$,
there are three situations: (A) $T_*<m_{\phi}<T_{RH}$, 
(B)  $T_*<T_{RH} \le m_{\phi}$, (C) $T_{RH}\le T_*<m_{\phi}$. As we
will see shortly, the evaporated charges in all cases are the same
order of magnitude. Therefore, we will show only the case (A) here. 
For $T>T_{RH}$,
\begin{equation}
    \frac{dQ}{dT} \sim 10 \frac{MT_{RH}^2}{T^4},
\end{equation}
so we have
\begin{eqnarray}
    \Delta Q(T>T_{RH}) & \equiv & Q_{init}-Q_{RH} \nonumber \\
      & \sim & 3MT_{RH}^2 \left(\frac{1}{T_{RH}^3}
              -\frac{1}{T_{init}^3}\right),
\end{eqnarray}
where subscript `init' denotes the initial values. If we assume
$T_{init} \gg T_{RH}$, we obtain $\Delta Q(T>T_{RH}) \sim 3M/T_{RH}$. 
For $T_*<T<T_{RH}$,
\begin{equation}
    \frac{dQ}{dT} \sim 10 \frac{M}{T^2},
\end{equation}
so the evaporated charge in this period is estimated as
\begin{equation}
    \Delta Q(T_*<T<T_{RH}) 
    \sim 10M  \left(\frac{1}{T_*}-\frac{1}{T_{RH}}\right).
\end{equation}
Finally, for $T<T_*$, we must integrate equation
\begin{equation}
    \frac{dQ}{dT} \sim 10 \frac{MT}{m_{\phi}^3}Q^{1/4}.
\end{equation}
We thus get 
\begin{equation}
    \Delta Q(T<T_*) \sim 5 \frac{MT_*^2}{m_{\phi}^3} Q_0^{1/4}, 
\end{equation}
where $Q_0$ is the amount of the Q-ball charge at present. Adding
three evaporated charges, we have
\begin{equation}
    \Delta Q \sim 10\frac{M}{T_*} - 7\frac{M}{T_{RH}}
         + 5 \frac{MT_*^2}{m_{\phi}^3} Q_0^{1/4}.
\end{equation}
If $T_{RH}\gg T_*$, the second term can be neglected. On the other
hand, if $T_{RH}\gtrsim T_*$, the sum of the first and second terms is
the same order of magnitude as the first one. Therefore, from the
first and second terms, we get
\begin{eqnarray}
    \label{first-second}
    \Delta Q^{(1+2)} & \sim & 10\frac{M}{T_*} 
      \sim 10\frac{M}{m_{\phi}}Q_{init}^{1/12} \nonumber \\
      & & \sim 2.4\times 10^{18} 
                  \left(\frac{m_{\phi}}{{\rm TeV}}\right)^{-1} 
                  \left(\frac{Q_{init}}{10^{24}}\right)^{1/12},
\end{eqnarray}
where we use Eq.(\ref{crit-T}) in the first line. For large enough
charge such as $Q_{init}\gg10^{18}$, there is little difference
between $Q_{init}$ and $Q_0$, so the third term can be estimated as
\begin{equation}
    \Delta Q^{(3)} 
      \sim 5\frac{M}{m_{\phi}^3}m_{\phi}^2 Q_{init}^{-1/6}Q_0^{1/4}
      \sim 5\frac{M}{m_{\phi}}Q_{init}^{1/12},
\end{equation}
which is the same order of magnitude as the sum of first and second
terms. On the other hand, if $Q_{init} < 10^{18}$, all the charges are 
evaporated before the temperature drops to $T_*$, and there is no
contribution from the temperature below $T_*$, which is the third term 
$\Delta Q^{(3)}$.

For the case (B), the reheating temperature is almost the same as
the transition temperature $T_*$, the amount of evaporated charge
should be about the same as Eq.(\ref{first-second}). Although it is not 
apparent in the case (C), we will now see that the amount of
evaporated charge is as the same order of magnitude as in the case
(A). Above the temperature $T_*$, the diffusion rate determines the
speed of the whole process, and it is exactly the same as the case
(A). Thus $\Delta Q(T>T_*)\sim3M/T_*$. For $T_{RH}<T<T_*$, the rate is 
determined by the evaporation rate, and we should use
\begin{equation}
    \frac{dQ}{dT} \sim 10 \frac{MT_{RH}^2}{m_{\phi}^3T}Q^{1/4}.
\end{equation}
Integrating this equation, we obtain
\begin{equation}
    \label{e-evap-1}
    \Delta Q(T_{RH}<T<T_*) \sim 10 
         \frac{MT_{RH}^2}{m_{\phi}^3}Q^{1/4} \log\frac{T_*}{T_{RH}}.
\end{equation}
On the other hand, for $T<T_{RH}$, we must integrate
\begin{equation}
    \frac{dQ}{dT} \sim 10 \frac{MT}{m_{\phi}^3}Q^{1/4},
\end{equation}
and the result is 
\begin{equation}
    \label{e-evap-2}
    \Delta Q(T<T_{RH}) \sim 10 \frac{MT_{RH}^2}{m_{\phi}^3}Q^{1/4}.
\end{equation}
Since $T_{RH} \lesssim T_*$, Equations (\ref{e-evap-1}) and
(\ref{e-evap-2}) have almost the same form, we can regard the charge
evaporation in these stages as
\begin{equation}
    \Delta Q \sim 10 \frac{MT_{RH}^2}{m_{\phi}^3}Q^{1/4}.
\end{equation}
For the contribution from this term to be dominant, we must have
condition $T_{RH} \gtrsim m_{\phi}Q^{-1/12}$. However, since 
$T_{RH}\lesssim T_* \sim m_{\phi}Q^{-1/12}$, contributions from this
stage is at most comparable to that in the region $T>T_*$. Therefore,
we can adopt the estimation of the evaporated charge from the Q ball
as 
\begin{equation}
    \label{evap-Q}
    \Delta Q \sim 10 \frac{M}{m_{\phi}}Q^{1/12}
       \sim 2.4\times 10^{18} 
              \left(\frac{m_{\phi}}{{\rm TeV}}\right)^{-1}
              \left(\frac{Q}{10^{24}}\right)^{1/12},
\end{equation}
for any cases.

\subsection{Gravity-mediation type Q balls}
Now we will show the evaporated charges for the `new' type of stable Q
ball \cite{KK3}. The evaporation and diffusion rate have the same
forms in terms of Q-ball parameters $R_Q$ and $\omega$. The only
differences are that we have to use the features for the
`gravity-mediation' type Q ball, such as 
\begin{equation}
    R_Q \sim |K|^{-1/2} m_{3/2}, \qquad \omega \sim m_{3/2},
\end{equation}
and the transition temperature when $\Gamma_{evap}=\Gamma_{diff}$
becomes $T_*\equiv A^{1/3}|K|^{1/6}(m_{3/2}m_{\phi}^2)^{1/3}$. As in
the `usual' type of Q balls, where the potential is dominated by the
logarithmic term, the charge evaporation near $T_*$ is dominant, and
the total evaporated charges are found to be \cite{KK3}
\begin{equation}
    \Delta Q \sim 10^{20} \left(\frac{m_{3/2}}{\rm MeV}\right)^{-1/3}
       \left(\frac{m_{\phi}}{\rm TeV}\right)^{-2/3}.
\end{equation}

\section{Cosmological Q-ball scenario}
\label{scenario}
\subsection{Usual gauge-mediation type Q balls}
Now we would like to see whether there is any consistent cosmological
scenario for the baryogenesis and the dark matter of the universe,
provided by large Q balls. In the first place, we will look for the
situation in which the logarithmic term dominates the AD potential,
and the `usual' gauge-mediation type of Q balls are formed. 

Speaking very loosely, we know that the amount of the baryons in the
universe is as large as that of the dark matter (within a few orders
of magnitude). In the Q-ball scenario, the baryon number of the
universe should be explained by the amount of the charge evaporated
from Q balls, $\Delta Q$, and the survived Q balls become the dark
matter. If we assume that Q balls do not exceed the critical density
of the universe, i.e., $\Omega_Q \lesssim 1$, and the baryon-to-photon
ratio as $\eta_B \sim 10^{-10}$, the condition can be written as 
\begin{equation}
    \label{eta-B}
    \eta_B = \frac{n_B}{n_{\gamma}} 
       \simeq \frac{\varepsilon n_Q\Delta Q}{n_{\gamma}}
       \simeq \frac{\varepsilon\rho_Q \Delta Q}{n_{\gamma}M_Q}
       \simeq \frac{\varepsilon\rho_{c,0} \Omega_Q \Delta Q}
                   {n_{\gamma,0}M_Q},
\end{equation}
where $\Omega_Q$ is the density parameter for the Q ball. Using 
$M_Q \simeq m_{\phi}Q^{3/4}$, 
$\rho_{c,0} \sim 8h_0^2\times 10^{-47} {\rm GeV}^4$, and
$n_{\gamma,0} \sim 3.3 \times 10^{-39} {\rm GeV}^3$, where 
$h_0(\sim 0.7)$ is the Hubble parameter normalized with 100km/sec/Mpc,
we obtain the ratio of the charges evaporated to be the baryons in the
universe and remaining in the dark matter Q ball: 
\begin{eqnarray}
    \label{dQ-Q-ratio}
    r_B \equiv \frac{\Delta Q}{Q} & &
      \sim \eta_B \frac{m_{\phi}n_{\gamma,0}}
           {\varepsilon\rho_{c,0}\Omega_Q} Q^{-1/4}
    \nonumber \\ & &
      \sim 10^{11} \varepsilon^{-1} \eta_B \Omega_Q^{-1}
            \left(\frac{m_{\phi}}{\rm TeV}\right) Q^{-1/4}.
\end{eqnarray}

As we mentioned earlier, in the inflaton-oscillation dominated
universe between inflation and reheating, the temperature has
different dependence on the time (or the Hubble parameter) from that
of the radiation dominated universe, and it reads as, from
Eq.(\ref{T-br}),
\begin{equation}
    \label{T-br-2}
    T \sim (M T_{RH}^2 H)^{1/4}.
\end{equation}
At the beginning of the AD field rotation, 
$H\sim m^2(T)/\phi_0 \sim T^2/\phi_0$. Inserting this into 
Eq.(\ref{T-br-2}) and rephrasing it, we obtain
\begin{equation}
    \label{Tr}
    T \sim T_{RH} \sqrt{\frac{M}{\phi_0}}.
\end{equation}

In the Affleck-Dine mechanism with Q-ball production, the baryon
number of the universe can be estimated as
\begin{equation}
    \label{eta-f}
    \eta_B \sim \frac{n_{B,RH}}{\rho_{I,RH}/T_{RH}}
           \sim \frac{n_{B,f}}{\rho_{I,f}/T_{RH}},
\end{equation}
where subscript $RH$ and $f$ denote the values at the time of the
reheating and the formation of Q balls, respectively. Notice that
$n_B$ and $\rho_I$ are proportional to $a^{-3}$ during the
inflaton-oscillation dominated universe before reheating. Thus we must
have the baryon number density at the formation time which will be
evaporated later. It can be written as
\begin{eqnarray}
    \label{n-b}
    n_{B,f} \sim r_B \varepsilon n_{\phi,f} 
    & \sim & r_B \varepsilon \omega \phi_0^2 \nonumber \\
    & \sim & 10^{11} \eta_B \Omega_Q^{-1}
      \left(\frac{m_{\phi}}{\rm TeV}\right) T^2 \phi_0 Q^{-1/4}, 
\end{eqnarray}
where we use Eq.(\ref{dQ-Q-ratio}), and 
$\omega \sim m^2(T)/\phi_0 \sim T^2/\phi_0$. On the other hand, the
energy density of the inflaton when the AD field starts its rotation
is 
\begin{equation}
    \label{rho-inf}
    \rho_{I,f} \sim H_{osc}^2M^2 \sim \frac{T^4}{\phi_0^2}M^2.
\end{equation}
Thus, from Eq.(\ref{eta-f}), we can write the baryon-to-photon
ratio $\eta_B$ as
\begin{equation}
    \eta_B \sim 
      10^{11} \eta_B \Omega_Q^{-1} 
        \left(\frac{m_{\phi}}{\rm TeV}\right)
        \frac{\phi_0^4}{T_{RH}M^3}Q^{-1/4},
\end{equation}
where Eqs.(\ref{Tr}), (\ref{n-b}), and (\ref{rho-inf}) are used. We
thus get the Q-ball charge as the function of the initial amplitude of
the AD field and the reheating temperature as
\begin{equation}
    \label{Q-phi-Trh}
    Q \sim 10^{44} 
      \left(\frac{m_{\phi}}{\rm TeV}\right)^4
      \frac{\phi_0^{16}}{T_{RH}^4M^{12}} \Omega_Q^{-4}.
\end{equation}

Now we will obtain another relation between the initial amplitude of
the AD field and the reheating temperature. From the analytical and
numerical estimation for the Q-ball charge, we have 
\begin{equation}
    \label{Q-simu}
    Q \sim \beta \left(\frac{\phi_0}{T}\right)^4,
\end{equation}
for $\varepsilon \sim 1$. For $\varepsilon \ll 1$, we have to replace
$\beta$ by $\beta'=\gamma\beta$ with $\gamma\sim 0.1$. 

Since we obtain two different expressions for the Q-ball charge, we
can get the initial amplitude of the AD field from
Eqs.(\ref{Q-phi-Trh}) and (\ref{Q-simu}) as, taking into account the
$\varepsilon$-dependence when $\varepsilon \sim 1$ in
Eq.(\ref{Q-simu}), 
\begin{eqnarray}
    \label{init-amp}
    \phi_0 & \sim & 4.6\times 10^{13} 
    \varepsilon^{1/10} \Omega_Q^{2/5}
    \nonumber \\ & & \qquad \times
    \left(\frac{\beta}{6\times 10^{-4}}\right)^{1/10}
    \left(\frac{m_{\phi}}{\rm TeV}\right)^{-2/5} {\rm GeV}.
\end{eqnarray}

Inserting this equation into Eq.(\ref{Tr}), we get the corresponding
temperature when the AD field starts the rotation. It reads as
\begin{eqnarray}
    T \sim & & 2.3\times 10^7 \varepsilon^{-1/20} \Omega_Q^{-1/5}
    \left(\frac{T_{RH}}{10^5 {\rm GeV}}\right) \nonumber \\
    & & \hspace*{5mm} \times
    \left(\frac{\beta}{6\times 10^{-4}}\right)^{-1/20}
    \left(\frac{m_{\phi}}{\rm TeV}\right)^{1/5} {\rm GeV}.
\end{eqnarray}
We also obtain the charge of the Q ball
\begin{eqnarray}
    \label{Q-limit}
    Q \sim 9.3 \times 10^{21} & & \varepsilon^{8/5} \Omega_Q^{12/5}
    \left(\frac{T_{RH}}{10^5 {\rm GeV}}\right)^{-4} \nonumber \\
    & &  \times
    \left(\frac{\beta}{6\times 10^{-4}}\right)^{8/5}
    \left(\frac{m_{\phi}}{\rm TeV}\right)^{-12/5},
\end{eqnarray}
where we use the numerical estimation (\ref{Q-simu}), and insert
Eqs.(\ref{Tr}) and (\ref{init-amp}) into it. Notice that the charge
$Q$ does not depend on the amount of the baryons, so that it just
represents the charge of the dark matter Q ball.

We must know how circular the orbit of the AD field motion is. It is
necessary to obtain the expression for $\varepsilon$ in terms of
other variables, say, the reheating temperature $T_{RH}$. In addition
to the condition of the amount of the evaporated charge for explaining
both the baryons and the dark matter simultaneously, which can be seen
in Eq.(\ref{dQ-Q-ratio}), we also have the estimation of the
evaporated charge from the Q ball (\ref{evap-Q}). Equating these two,
we have 
\begin{equation}
    \label{dm-Q}
    Q \sim 1.2 \times 10^8 \varepsilon^{3/2}\eta_B^{-3/2}
    \Omega_Q^{3/2}\left(\frac{m_{\phi}}{\rm TeV}\right)^{-3}.
\end{equation}
Compared with Eq.(\ref{Q-limit}), $\varepsilon$ should be
\begin{eqnarray}
    \varepsilon \sim 1.3\times10^{11} \Omega_Q^{-9} & &
    \left(\frac{\eta_B}{10^{-10}}\right)^{-15}
    \left(\frac{T_{RH}}{10^5 {\rm GeV}}\right)^{40}
    \nonumber \\ & & \times
    \left(\frac{\beta}{6\times 10^{-4}}\right)^{-16}
    \left(\frac{m_{\phi}}{\rm TeV}\right)^{-6}.
\end{eqnarray}
for explaining the amount of the baryons and the dark matter of the
universe simultaneously. In the case of $\varepsilon \ll 1$,
Eq.(\ref{Q-simu}) has no $\varepsilon$-dependence, so
Eqs.(\ref{init-amp}) $-$ (\ref{Q-limit}) also have no
$\varepsilon$-dependence. Thus, we instead obtain the formula for
$\varepsilon$ as
\begin{eqnarray}
    \varepsilon \sim 1.5 \times 10^{-2} \Omega_Q^{3/5} & &
    \left(\frac{\eta_B}{10^{-10}}\right)
    \left(\frac{T_{RH}}{10^5 {\rm GeV}}\right)^{-8/3}
    \nonumber \\ & & \times
    \left(\frac{\beta}{6\times 10^{-5}}\right)^{16/15}
    \left(\frac{m_{\phi}}{\rm TeV}\right)^{2/5}.
\end{eqnarray}

To summarize, the conditions of parameters for the Q-ball baryogenesis
and dark matter scenario to work are given as
\begin{eqnarray}
        \label{sum-phi}
    \phi_0 & \sim & 4.6 \times 10^{13} 
    \varepsilon^{1/10}\Omega_Q^{2/5} \nonumber \\
    & & \hspace*{10mm} \times
    \left(\frac{\beta}{6\times 10^{-4}}\right)^{1/10} 
    \left(\frac{m_{\phi}}{\rm TeV}\right)^{-2/5} {\rm GeV},  \\
        \label{sum-T}
    T & \sim & 2.3\times 10^7 \varepsilon^{-1/20} \Omega_Q^{-1/5}
    \left(\frac{T_{RH}}{10^5 {\rm GeV}}\right) \nonumber \\
    & & \hspace*{10mm} \times
    \left(\frac{\beta}{6\times 10^{-4}}\right)^{-1/20}
    \left(\frac{m_{\phi}}{\rm TeV}\right)^{1/5} {\rm GeV},
    \\
        \label{sum-Q}
    Q & \sim & 9.3 \times 10^{21} \varepsilon^{8/5} \Omega_Q^{12/5}
    \left(\frac{T_{RH}}{10^5 {\rm GeV}}\right)^{-4} \nonumber \\
    & & \hspace*{15mm} \times
    \left(\frac{\beta}{6\times 10^{-4}}\right)^{8/5}
    \left(\frac{m_{\phi}}{\rm TeV}\right)^{-12/5},
    \\
        \label{sum-eps}
    \varepsilon & \sim & 1.3\times10^{11} \Omega_Q^{-9}
    \left(\frac{\eta_B}{10^{-10}}\right)^{-15}
    \left(\frac{T_{RH}}{10^5 {\rm GeV}}\right)^{40}
    \nonumber \\ & & \hspace*{17mm} \times
    \left(\frac{\beta}{6\times 10^{-4}}\right)^{-16}
    \left(\frac{m_{\phi}}{\rm TeV}\right)^{-6},
\end{eqnarray}
where we should omit $\varepsilon$-dependences in Eqs.(\ref{sum-phi}) 
$-$ (\ref{sum-Q}) for $\varepsilon \ll 1$.

As we mentioned in Sec.~\ref{AD-mech},  the initial amplitude of
the AD field is determined by the balance between the Hubble mass 
term and the nonrenormalizable term. When the AD field starts rolling
down its potential, the amplitude becomes 
$\phi_0 \sim (H_{osc}M^{n-3})^{1/(n-2)} $ for the superpotential
$W\sim\phi^n/M^{n-3}$, where $H_{osc}\sim T^2/\phi_0$. Using
Eq.(\ref{Tr}), we can write it as
\begin{equation}
    \phi_0 \sim (T_{RH}^2 M^{n-2})^{1/n}.
\end{equation}

We will see the range of $n$ in which we can obtain the consistent
scenario naturally. At the first place, let us consider the case 
$n=4$, where $\varepsilon \ll 1$ will be led below. In this
case, the initial amplitude becomes 
\begin{equation}
    \phi_0 \sim 4.9\times 10^{11}
    \left(\frac{T_{RH}}{10^5 {\rm GeV}}\right)^{1/2} {\rm GeV}.
\end{equation}
Therefore, the required reheating temperature should be
\begin{equation}
    T_{RH} \sim 8.8\times 10^8 \Omega_Q^{4/5}
           \left(\frac{\beta}{6\times10^{-4}}\right)^{1/5}
           \left(\frac{m_{\phi}}{\rm TeV}\right)^{-4/5} {\rm GeV}, 
\end{equation}
where Eq.(\ref{sum-phi}) is used. Then, for this reheating 
temperature, we have
\begin{eqnarray}
    T & \sim & 2.0 \times 10^{11} \Omega_Q^{3/5}
    \nonumber \\ & & \quad \times
    \left(\frac{\beta}{6\times10^{-4}}\right)^{3/20}
    \left(\frac{m_{\phi}}{\rm TeV}\right)^{-3/5} {\rm GeV}, \\ 
    Q & \sim & 1.6\times 10^6 \Omega_Q^{-4/5}
    \left(\frac{\beta}{6\times10^{-4}}\right)^{4/5}
    \left(\frac{m_{\phi}}{\rm TeV}\right)^{4/5}, \\
    \varepsilon & \sim & 8.8\times 10^{-8} \Omega_Q^{-23/15}
    \left(\frac{\eta_B}{10^{-10}}\right)
    \nonumber \\ & & \hspace{10mm} \times
    \left(\frac{\beta}{6\times10^{-4}}\right)^{8/15}
    \left(\frac{m_{\phi}}{\rm TeV}\right)^{38/15},
\end{eqnarray}
so that all the charges evaporate and no dark matter Q balls
exists. For the $n=5$ case, 
\begin{equation}
    \phi_0 \sim 1.1\times 10^{13} 
        \left(\frac{T_{RH}}{10^5 {\rm GeV}}\right)^{2/5} 
                        {\rm GeV},
\end{equation}
and the reheating temperature should be
\begin{equation}
    T_{RH} \sim 3.6\times 10^6 \Omega_Q
                \left(\frac{\beta}{6\times10^{-4}}\right)^{1/4}
                \left(\frac{m_{\phi}}{\rm TeV}\right)^{-1} {\rm GeV},
\end{equation}
which leads to, taking into account the fact that $\varepsilon \ll 1$
in this case,
\begin{eqnarray}
    & & T \sim 8.3\times 10^8 \Omega_Q^{4/5}
        \left(\frac{\beta}{6\times10^{-4}}\right)^{1/5}\!\!
        \left(\frac{m_{\phi}}{\rm TeV}\right)^{-4/5} {\rm GeV}, \\
    & & Q \sim 5.5\times 10^{15} \Omega_Q^{-8/5}
        \left(\frac{\beta}{6\times10^{-4}}\right)^{3/5}
        \left(\frac{m_{\phi}}{\rm TeV}\right)^{8/5}, \\
    & & \varepsilon \sim 1.3\times 10^{-5} \Omega_Q^{-31/15}
        \left(\frac{\eta_B}{10^{-10}}\right)
        \nonumber \\ & & \hspace*{30mm} \times
        \left(\frac{\beta}{6\times10^{-4}}\right)^{2/5}
        \left(\frac{m_{\phi}}{\rm TeV}\right)^{46/15}. 
\end{eqnarray}
In order for the Q ball to survive from the evaporation, the initial 
charge of the Q ball should be large enough. This condition is
$Q \gtrsim \Delta Q$, and can be achieved from Eq.(\ref{evap-Q}) if
\begin{equation}
    \label{survive-2}
    Q \gtrsim 7.4\times 10^{17} 
      \left(\frac{m_{\phi}}{\rm TeV}\right)^{-12/11}. 
\end{equation}
Imposing this condition on the required Q-ball charge above, we obtain
the required mass of the AD particle as
\begin{equation}
    m_{\phi} \gtrsim 6.2\times 10^3 \Omega_Q^{22/37}
      \left(\frac{\beta}{6\times10^{-4}}\right)^{-33/148} {\rm GeV},
\end{equation}
so that the degree of the ellipticity of the orbit of the AD field
motion should be
\begin{equation}
    \varepsilon \gtrsim 3.5 \times 10^{-3} \Omega_Q^{-\frac{533}{111}} 
        \left(\frac{\eta_B}{10^{-10}}\right)
        \left(\frac{\beta}{6\times10^{-4}}\right)^{-21/74}.
\end{equation}
Since, $\varepsilon \ll 1$ in this case, we have to use an order of
magnitude smaller value for $\beta$. Then, we get the following values
for the parameters in order for the Q-ball scenario to work naturally:
\begin{eqnarray}
    \label{good-5}
    m_{\phi} & \sim & 1.0\times 10^4 \beta_s^{-33/148} 
        \Omega_Q^{22/37} \ {\rm GeV}, \nonumber \\[2mm]
    \varepsilon & \sim & 6.7\times 10^{-3} \eta_{B,10} 
        \beta_s^{-21/74} \Omega_Q^{-533/111}, \nonumber \\[2mm]
    T & \sim & 8.3 \times 10^7 \beta_s^{14/37}
        \Omega_Q^{12/37} \ {\rm GeV}, \nonumber \\[2mm]
    T_{RH} & \sim & 2.0\times 10^5  \beta_s^{35/74}
        \Omega_Q^{15/37} \ {\rm GeV}, \nonumber \\[2mm]
    \phi_0 & \sim & 1.5 \times 10^{13} \beta_s^{7/37}
        \Omega_Q^{6/37} \ {\rm GeV}, \nonumber \\[2mm]
    Q & \sim & 5.5 \times 10^{16} \beta_s^{9/37}
        \Omega_Q^{-24/37},
\end{eqnarray}
where $\beta_s\equiv \beta/(6\times 10^{-5})$ and 
$\eta_{B,10}\equiv \eta_B/10^{-10}$. Actually, this parameter set is
the lower limit for $\varepsilon$, and larger values are also
allowed. As we see later, the upper bound comes from the condition
that the Q ball is stable against the decay into nucleons. The allowed
range is $6.7\times 10^{-3} < \varepsilon < 5.1\times 10^{-2}$, or,
equivalently, $1.0\times 10^4 {\rm GeV} < m_{\phi} < 2.0\times 10^4$
GeV.

Now let us move on to the $n=6$ case. Repeating the similar argument,
we obtain the consistent values for parameters
\begin{eqnarray}
    \label{good-6}
    \varepsilon & \sim & 0.97 \Omega_Q^{-39/11} \eta_{B,10}^{15/11}
       \beta_{\ell}^{4/11} m_{\phi,430}^{54/11}, \nonumber \\[2mm]
    T & \sim & 9.1 \times 10^6 \varepsilon^{1/4} \Omega_Q 
       \beta_{\ell}^{1/4} m_{\phi,430}^{-1} \ {\rm GeV}, 
       \nonumber \\[2mm]
    T_{RH} & \sim & 4.7\times 10^4 \varepsilon^{3/10} \Omega_Q^{6/5} 
       \beta_{\ell}^{3/10} m_{\phi,430}^{-6/5} \ {\rm GeV}, 
       \nonumber \\[2mm]
    \phi_0 & \sim & 6.4\times 10^{13}\varepsilon^{1/10}\Omega_Q^{2/5} 
       \beta_{\ell}^{1/10} m_{\phi,430}^{-2/5} \ {\rm GeV}, 
       \nonumber \\[2mm] 
    Q & \sim & 1.5 \times 10^{24} \varepsilon^{2/5} \Omega_Q^{-12/5} 
       \beta_{\ell}^{2/5} m_{\phi,430}^{12/5},
\end{eqnarray}
where $\beta_{\ell}\equiv \beta/(6\times10^{-4})$ and 
$m_{\phi,430}\equiv m_{\phi}/$(430 GeV). In this case, the AD field
rotates circularly and produce the baryon number maximally. In
general, however, $\varepsilon \ll 1$ cases are also allowed. 
Numerical calculations reveal that the $m_{\phi}$-dependence of the
Q-ball charge changes around $\varepsilon \sim 0.1$ using the
following results shown in the previous section: $Q$ is proportional
to $\varepsilon$ for $\varepsilon \gtrsim 0.1$, while constant for
smaller values, which reflects the production of both positive and
negative charged Q balls. Therefore, the charge can be written as
\begin{equation}
    Q \propto \left\{
      \begin{array}{ll}
          m_{\phi}^{12/5} & (\varepsilon \lesssim 0.1), \\[2mm]
          m_{\phi}^{48/11} & (\varepsilon \gtrsim 0.1), \\
      \end{array}
      \right.
\end{equation}
and the lower bound may be determined by the possible lowest mass,
$m_{\phi} \gtrsim 100$ GeV.

For the $n=7$ case, we need extremely huge $\varepsilon$ such as 
$\sim 6\times 10^8$, which cannot be realized. Higher $n$ makes the
situation worse. 

In order for the scenario to work naturally, we must check that the
values of $\varepsilon$ are consistent with the A-terms derived from
the nonrenormalizable superpotential. As we derived in
Sec.~\ref{AD-mech}, the formula for the $\varepsilon$ is written as
\begin{equation}
    \varepsilon \sim \frac{m_{3/2}\phi_0}{T^2}.
\end{equation}

For $n=5$ case, putting the values for $\phi_0$ and $T$ in
Eq.(\ref{good-5}), we have 
\begin{equation}
    \varepsilon \sim 2.2\times 10^{-3} \Omega_Q^{-18/37}
      \left(\frac{\beta}{6\times10^{-5}}\right)^{-21/37}
      \left(\frac{m_{3/2}}{\rm GeV}\right). 
\end{equation}
Comparing this with the value of $\varepsilon$ in Eq.(\ref{good-5}), 
we get
\begin{equation}
    m_{3/2} \sim 0.33 \Omega_Q^{-479/111}
      \left(\frac{\beta}{6\times10^{-5}}\right)^{-21/74}
      \left(\frac{\eta_B}{10^{-10}}\right) {\rm GeV}. 
\end{equation}
Therefore, we get a consistent Q-ball scenario naturally in the $n=5$
case. Notice that the reheating temperature is low enough for
$m_{3/2}\sim 1$ GeV to avoid the cosmological gravitino problem
\cite{Moroi}. Notice that, for the larger $\varepsilon$ cases,
$m_{3/2}$ gets too large in the framework of the gauge-mediated SUSY
breaking, and the allowed range becomes $m_{\phi}\simeq 10 - 14$ TeV. 

In the $n=6$ case, we have
\begin{eqnarray}
    \varepsilon & \sim & 0.83 \Omega_Q^{-8/7}
      \left(\frac{\beta}{6\times10^{-4}}\right)^{-2/7}
      \nonumber \\ & & \hspace*{10mm} \times
      \left(\frac{m_{\phi}}{430 {\rm GeV}}\right)^{8/7}
      \left(\frac{m_{3/2}}{\rm GeV}\right)^{5/7},
\end{eqnarray}
for the values of $\phi_0$ and $T$ in Eq.(\ref{good-6}), and comparing 
with the value of $\varepsilon$ in Eq.(\ref{good-6}), we get
\begin{eqnarray}
    m_{3/2} & \sim &1.2 \ \Omega_Q^{-51/21} 
      \left(\frac{\beta}{6\times10^{-4}}\right)^{10/11}
      \nonumber \\ & &  \times
      \left(\frac{m_{\phi}}{430 {\rm GeV}}\right)^{297/55}
      \left(\frac{\eta_B}{10^{-10}}\right)^{21/11} {\rm GeV}.
\end{eqnarray}
Therefore, we obtain the consistent scenario for $n=6$, which is
achieved, for example, if we choose the $udd$ flat direction for the
AD field and use $W \sim (udd)^2$. Notice again that no gravitino
problem exist in this scenario. The scenario also works for smaller
$\varepsilon$ for $m_{3/2} \sim 1$ GeV, and the allowed range of the
parameter is $2.8 \times 10^{-3} \lesssim \varepsilon \lesssim 1$, or,
equivalently, $100 {\rm GeV} \lesssim m_{\phi} \lesssim 430$ GeV.

Now we must see whether the effective potential of the AD field is
really dominated by the logarithmic term over the gravity-mediation
term at this field amplitude. This condition holds if
\begin{equation}
    T^4 \log \left( \frac{\phi_0^2}{T^2}\right)
    \gtrsim m_{3/2}^2 \phi_0^2,
\end{equation}
or, equivalently,
\begin{equation}
    T \gtrsim (m_{3/2}\phi_0)^{1/2}
    \left[ \log \left( \frac{\phi_0^2}{T^2}\right)\right]^{-1/4}.
\end{equation}
If we use the values in Eqs.(\ref{good-5}) and (\ref{good-6}), the
right hand sides becomes $\sim 1.0\times 10^6$ GeV and 
$\sim 3.8\times 10^6$ GeV, respectively. Thus, the condition is met,
and we have consistent cosmological scenarios in the gauge-mediated
SUSY breaking model for the effective potential dominated by the
(thermal) logarithmic term. 

\subsection{Gravity-mediation type Q balls}
For the thorough investigation, we should consider whether the `new'
type Q-ball scenario works for low enough reheating temperature
avoiding the thermal effects. We will follow the same argument as we
did for the `usual' gauge-mediation type Q ball. From the
baryon-to-photon ratio (\ref{eta-B}), the ratio of the charge
evaporated from the Q ball to that remained in the Q ball can be
estimated as
\begin{equation}
    r_B \sim \frac{\Delta Q}{Q} \sim 10^5 \eta_B \Omega_Q^{-1} 
                             \left(\frac{m_{3/2}}{\rm MeV}\right),
\end{equation}
where we have used $M_Q \sim m_{3/2}Q$ and put $\varepsilon=1$, since
it is a natural realization (see Sec.~\ref{AD-mech}). Since the
baryon-to-photon ratio can also be written as  
\begin{eqnarray}
    \eta_B & \sim & \frac{n_{B,osc}}{\rho_{I,osc}/T_{RH}}
             \sim \frac{r_B m_{3/2}\phi_0^2}
                     {m_{3/2}^2M^2/T_{RH}} \nonumber \\
           & & \sim 10^{13} \eta_B \Omega_Q^{-1}
             \left(\frac{T_{RH}}{10^5 {\rm GeV}}\right)
                \left(\frac{\phi_0}{M}\right)^2,
\end{eqnarray}
the reheating temperature can be estimated as
\begin{equation}
    \label{Trh-new}
    T_{RH} \sim 10^{-8} \ \Omega_Q
        \left(\frac{\phi_0}{M}\right)^{-2} {\rm GeV}.
\end{equation}

In the previous section, we have obtained the relation between the
formed Q-ball charge and the initial amplitude of the AD field as
\begin{equation}
    \label{Q-simu-new}
    Q = \tilde{\beta}\left(\frac{\phi_0}{m_{3/2}}\right)^2,
\end{equation}
where $\tilde{\beta}\approx 6\times 10^{-3}$ from our simulations. On
the other hand, we have the condition that the evaporated charge and
survived stable Q balls explain for the baryon and the dark matter of
the universe, respectively, as \cite{KK3}
\begin{equation}
    \label{Q-dm-new}
    Q \sim 10^{25} \Omega_Q 
      \left(\frac{m_{3/2}}{\rm MeV}\right)^{-4/3} 
      \left(\frac{m_{\phi}}{\rm TeV}\right)^{-2/3}.
\end{equation}
Therefore, from Eqs.(\ref{Q-simu-new}) and (\ref{Q-dm-new}), we obtain
the required amplitude of the AD field as
\begin{eqnarray}
    \phi_0 \sim 4\times 10^{10} & &\Omega_Q^{1/2} 
      \left(\frac{\tilde{\beta}}{6\times 10^{-3}}\right)^{-1/2}
      \nonumber \\ & & \times 
      \left(\frac{m_{3/2}}{\rm MeV}\right)^{1/3}
      \left(\frac{m_{\phi}}{\rm TeV}\right)^{-1/3} {\rm GeV}.
\end{eqnarray}
Inserting this into Eq.(\ref{Trh-new}), the required reheating
temperature is
\begin{eqnarray}
    \label{DM-detect}
    T_{RH} \sim 3.5\times 10^7 & & 
      \left(\frac{\tilde{\beta}}{6\times 10^{-3}}\right)
      \nonumber \\ & & \times
      \left(\frac{m_{3/2}}{\rm MeV}\right)^{-2/3}
      \left(\frac{m_{\phi}}{\rm TeV}\right)^{2/3} {\rm GeV}.
\end{eqnarray}

When the gravity-mediation term $m_{3/2}^2\phi^2$ dominates the AD
potential, the field starts its rotation at $H\sim m_{3/2}$, so the
temperature at that time is $T_{osc}\sim(M T_{RH}^2 m_{3/2})^{1/4}$. 
Thus we have
\begin{eqnarray}
    T_{osc} \sim 4 \times 10^7 & &
      \left(\frac{\tilde{\beta}}{6\times 10^{-3}}\right)^{1/2}
      \nonumber \\ & & \times
      \left(\frac{m_{3/2}}{\rm MeV}\right)^{-1/12}
      \left(\frac{m_{\phi}}{\rm TeV}\right)^{1/3} {\rm GeV}.
\end{eqnarray}
We must obtain the constraint that the gravity-mediation term
dominates over the thermal logarithmic term in order for the scenario
of the `new' type Q ball to be successful. It reads as 
\begin{equation}
    T_{osc}^4 \log \left(\frac{\phi_0^2}{T_{osc}^2}\right)
      \lesssim m_{3/2}^2\phi_0^2,
\end{equation}
and we can rephrase it as
\begin{equation}
    m_{3/2} \gtrsim 2.8\times 10^2 \Omega_Q^{-1/3}
      \left(\frac{\tilde{\beta}}{6\times 10^{-3}}\right)^{1/3}
      \left(\frac{m_{\phi}}{\rm TeV}\right)^{2/3} {\rm GeV}.
\end{equation}
Therefore, this range of the gravitino mass is too large for the
gauge-mediated SUSY breaking model, and the thermal logarithmic term
dominates the potential at these field values. We thus find no
consistent model for the 'new' type of stable Q balls.

\section{Generic models for gauge mediation}
\label{evap}
\subsection{Dominated by zero-temperature potential}
As we mentioned earlier, the scale of the logarithmic potential could
be much larger than $m_{\phi}$ in the general context of the
gauge-mediated SUSY breaking. For the complete consideration, we will
discuss this situation in this section. Here we will express it as
\begin{equation}
    \label{pot-app}
    V \sim \left\{
      \begin{array}{ll}
          \ds{M_F^4 \log \left(\frac{\phi^2}{M_S^2}\right)} 
          & (\phi \gg M_S), \\
          m_{\phi}^2 \phi^2 & (\phi \ll M_S), \\
      \end{array}
      \right.
\end{equation}
where $M_S$ is the messenger mass scale. Formation of the dark matter
Q balls will take place at large field amplitudes, so the mass, the 
size, etc. have to be different, and their expressions are as
follows:
\begin{equation}
    M_Q \sim M_F Q^{3/4}, \quad 
    R \sim M_F^{-1}Q^{1/4}, \quad
    \omega \sim \frac{M_F^2}{\phi}, \dots
\end{equation}
In order for this potential to dominate over the thermal logarithmic
potential, we need condition $M_F \gtrsim T$. Otherwise, the results
discussed in the next subsections have to  be applied. Since this
type of the Q ball should be stable against the decay into nucleons,
the Q-ball mass per unit charge must be smaller than 1 GeV. This
condition holds for
\begin{equation}
    \label{stable-app}
    Q \gtrsim 10^{24} \left(\frac{M_F}{10^6 {\rm GeV}}\right)^4.
\end{equation}

We can regard that all variables are rescaled with respect to $M_F$ in 
our simulations now. So the charge of the produced Q ball is
\begin{equation}
    \label{Q-simu-app}
    Q = \beta \left(\frac{\phi_0}{M_F}\right)^4,
\end{equation}
where $\beta\approx 6\times 10^{-4}$ again. The evaporation rate has
to be changed to
\begin{equation}
    \Gamma_{evap} = \frac{dQ}{dt} \sim \left\{
      \begin{array}{ll}
          \ds{-4\pi\frac{T^2}{M_F}Q^{1/4}} & (T>m_{\phi}), \\[2mm]
          \ds{-4\pi\frac{T^4}{m_{\phi}^2M_F}Q^{1/4}}&(T<m_{\phi}).\\  
      \end{array}
      \right.
\end{equation}
Depending on the Q-ball charge, the transition temperature $T_*$ where 
the diffusion and the evaporation rates are equal can be written as
\begin{equation}
    T_* \sim \left\{
      \begin{array}{ll}
          A M_F Q^{-1/4} & (T_*>m_{\phi}), \\[2mm]
          A^{1/3} (m_{\phi}^2M_F)^{1/3}Q^{-1/12}  & (T_*<m_{\phi}), \\
      \end{array}
      \right.
\end{equation}
where $A\sim 4$ is defined in Eq.(\ref{diff-rate}). These two
temperatures coincide when $Q=Q_{cr}\sim A^4 M_F^4 m_{\phi}^{-4}$.
Taking the stability condition (\ref{stable-app}) into account, we
must consider only the case with $Q>Q_{cr}$, which corresponds to
$T_*<m_{\phi}$ case. 

The charge transfer from inside the Q ball to its outside is
determined by the diffusion rate when $T>T_*$, so the evaporated
charge during this period is
\begin{eqnarray}
    & &\Delta Q(T>T_*)  \sim  10 \frac{M}{T_*} \nonumber \\
    & &\sim  4.6\times 10^{15} 
    \left(\frac{m_{\phi}}{10^2 {\rm GeV}}\right)^{-2/3}
    \left(\frac{M_F}{10^6 {\rm GeV}}\right)^{-1/3} Q^{1/12}.
\end{eqnarray}
On the other hand, when $T<T_*$, the charge transfer is determined by
the evaporation rate, and the evaporated charge can be estimated as
\begin{eqnarray}
    & & \Delta Q(T<T_*) \sim 
    \frac{MT_*^2}{m_{\phi}^2M_F}Q^{1/4}, \nonumber \\
    & & \sim 4.6\times 10^{15} 
    \left(\frac{m_{\phi}}{10^2 {\rm GeV}}\right)^{-2/3}
    \left(\frac{M_F}{10^6 {\rm GeV}}\right)^{-1/3} Q^{1/12}.
\end{eqnarray}
Therefore, the total charge evaporated from the Q ball can be written
as
\begin{equation}
    \label{charge-app}
    \Delta Q
    \sim 4.6\times 10^{15} 
    \left(\frac{m_{\phi}}{10^2 {\rm GeV}}\right)^{-2/3}
    \left(\frac{M_F}{10^6 {\rm GeV}}\right)^{-1/3} Q^{1/12}.
\end{equation}

Now we can impose the survival condition, which implies that the Q
ball should survive from the charge evaporation and become the dark
matter of the universe. It reads as $Q\gtrsim\Delta Q$, and equivalent
to 
\begin{equation}
    \label{survive-app}
    Q \gtrsim 1.2 \times 10^{17} 
    \left(\frac{m_{\phi}}{10^2 {\rm GeV}}\right)^{-8/11}
    \left(\frac{M_F}{10^6 {\rm GeV}}\right)^{-4/11}.
\end{equation}

Since the baryon number and the amount of the dark matter is related
as in Eq.(\ref{eta-B}), we get the charge of the Q balls for this type
as 
\begin{eqnarray}
    \label{B-DM-app}
    Q \sim 3.2 \times & 10^{17} &\Omega_Q^{3/2} \varepsilon^{3/2}
        \left(\frac{\eta_B}{10^{-10}}\right)^{-3/2} 
        \nonumber \\ & & \times
        \left(\frac{m_{\phi}}{10^2 {\rm GeV}}\right)^{-1}
        \left(\frac{M_F}{10^6 {\rm GeV}}\right)^{-2}.
\end{eqnarray}
We call this the baryon-dark matter (B-DM) condition.

In addition to three constraints mentioned above, there is another
one. The potential (\ref{pot-app}) at large field values must dominate 
over the thermal logarithmic potential, and this condition is set by
$M_F \gtrsim T$, as we mentioned above. Rewriting it, we obtain
\begin{eqnarray}
    M_F \gtrsim 2.7 & \times & 10^7 \varepsilon^{-5/12} \Omega_Q^{1/36} 
       \left(\frac{\beta}{6\times 10^{-4}}\right)^{7/18} \nonumber \\
       & & \times
       \left(\frac{\eta_B}{10^{-10}}\right)^{5/12}
       \left(\frac{m_{\phi}}{10^2 {\rm GeV}}\right)^{5/18} {\rm GeV},
\end{eqnarray}
where we have used the required initial amplitude of the AD field,
\begin{eqnarray}
    \phi_0 \sim 1.5 & \times & 10^{11}
        \varepsilon^{3/8}\Omega_Q^{1/36}  
        \left(\frac{\beta}{6\times 10^{-4}}\right)^{-1/4}
        \left(\frac{\eta_B}{10^{-10}}\right)^{-3/8} \nonumber \\
        & & \times
        \left(\frac{m_{\phi}}{10^2 {\rm GeV}}\right)^{-1/4}
        \left(\frac{M_F}{10^6 {\rm GeV}}\right)^{1/2} {\rm GeV},
\end{eqnarray}
which is derived by equating Eqs.(\ref{Q-simu-app}) and
(\ref{B-DM-app}).

\begin{figure}[t!]
\centering
\hspace*{-7mm}
\leavevmode\epsfysize=10cm \epsfbox{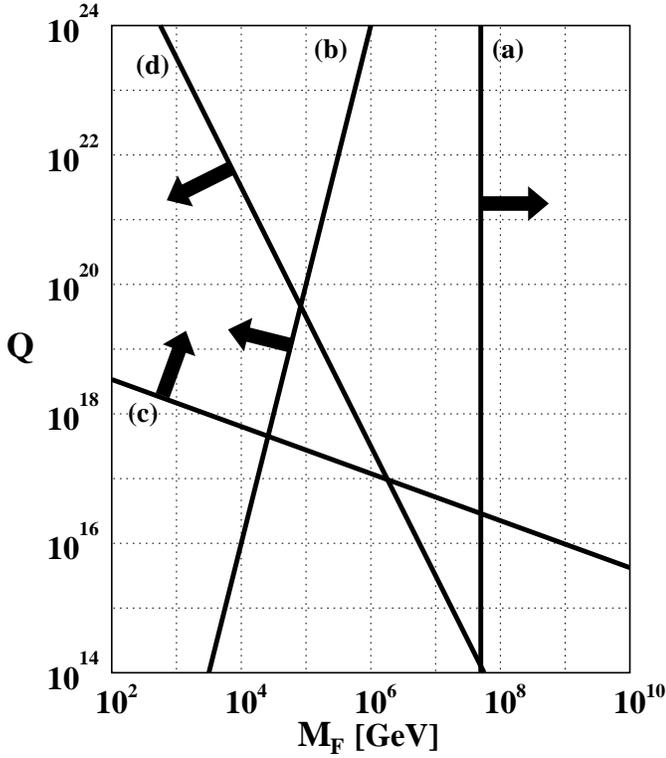}\\[2mm]
%\vspace{-3mm}
\caption[fig11]{\label{fig11} 
Constraints on ($Q,M_F$) plain with $m_{\phi}=100$ GeV. There are four
conditions: (a) non-dominant thermal logarithmic potential, (b)
stability against the decay into nucleons, (c) survival from charge
evaporation, and (d) right relation between the amounts of the baryon
and the dark matter.} 
\end{figure}

We can now put these four constraints on the parameter space
$(Q,M_F)$ as shown in Fig.~\ref{fig11}. In this figure, there are four 
constraints: line (a) is the condition that the potential is dominated
by Eq.(\ref{pot-app}), and the thermal logarithmic potential is
negligible, line (b) is the stability condition where the Q ball
cannot decay into nucleons and becomes the dark matter of the
universe, line (c) represents the survival condition that the Q ball
should survive from charge evaporation, and line (d) denotes the
B-DM condition that have the right relation between the amounts of the
baryons and the dark matter. We also show arrows to notice that which
side of these line are allowed. As can be seen, there is no allowed
region in the parameter space if we combine four conditions above,
which implies that this type of the Q ball is very difficult to
explain both the amount of the baryons and the dark matter
simultaneously.

\subsection{Dominated by finite-temperature potential}
Here we will investigate for the Q-ball formation in the generic
potential, but the thermal logarithmic term dominates the potential at 
the formation time. We have considered the special case $M_F=m_{\phi}$ 
in the previous section, and the results which we have derived above
can be used if we replace $m_{\phi}$ by $M_F$ appropriately. As we
mentioned in the last subsection, the only exception where the simple
replacing cannot be applied is the estimation of the evaporated
charges [see Eq.(\ref{charge-app})]. Similar analysis reveals that the 
formation of the Q balls takes place with the following parameters:
\begin{eqnarray}
    \label{init-phi-gen}
    \phi_0 & \sim & 2.9\times 10^{12} \varepsilon^{1/10}
    \Omega_Q^{2/5} \beta_{\ell}^{1/10} M_{F,6}^{-2/5} \ {\rm GeV}, \\
    \label{T-gen}
    T & \sim & 9.2\times 10^7 \varepsilon^{-1/20} \Omega_Q^{-1/5}
    \beta_{\ell}^{-1/20} M_{F,6}^{1/5} T_{RH,5} \ {\rm GeV}, \\
    Q & \sim & 5.9\times 10^{14} \Omega_Q^{12/5}\beta_{\ell}^{8/5}
    M_{F,6}^{-12/5} T_{RH,5}^{-4}, \\
    \varepsilon & \sim & 2.2\times 10^{27} \Omega_Q^{-9}
    \beta_{\ell}^{-16} \eta_{B,10}^{-15} m_{\phi,2}^{-10}
    M_{F,6}^{4} T_{RH,5}^{40},
\end{eqnarray}
where $\beta_{\ell} \equiv \beta/(6\times 10^{-4})$, 
$M_{F,6}\equiv M_F/(10^6 \ {\rm GeV})$,
$T_{RH,5}\equiv T_{RH}/(10^5 \ {\rm GeV})$,  
$\eta_{B,10}\equiv \eta_B/10^{-10}$, and
$m_{\phi,2}\equiv m_{\phi}/(10^2 {\rm GeV})$.

The initial amplitude of the field is determined by the balance
between the Hubble mass and nonrenormalizable terms. Since the
scenario naturally works only for $n=6$ case, we depict the results
for this case only. Since the initial value of the field is
\begin{equation}
    \phi_0 \sim (T_{RH}^2 M^4)^{1/6} 
    \sim 8.3\times 10^{13} T_{RH,5}^{1/3} \ {\rm GeV},
\end{equation}
we get the relation between the reheating temperature and $M_F$ as
\begin{equation}
    \label{T-M-rela}
    T_{RH,5} \sim 4.3\times 10^{-5} \varepsilon^{3/10} 
    \Omega_Q^{6/5} \beta_{\ell}^{3/10} M_{F,6}^{-6/5},
\end{equation}
where we have used Eq.(\ref{init-phi-gen}). Then we have 
\begin{eqnarray}
    \label{T-gen2}
    T & \sim & 4.0\times 10^3 \varepsilon^{1/4}\Omega_Q 
    \beta_{\ell}^{1/4} M_{F,6}^{-1} \ {\rm GeV}, \\
    \label{Q-gen}
    Q & \sim & 1.7\times 10^{32} \varepsilon^{3/5} 
    \Omega_Q^{-12/5} \beta_{\ell}^{2/5} M_{F,6}^{12/5}, \\
    \varepsilon & \sim & 2.5\times 10^{13} \Omega_Q^{-39/11} 
    \beta_{\ell}^{4/11} \eta_{B,10}^{15/11} M_{F,6}^4
    m_{\phi,2}^{10/11}.
\end{eqnarray}
For $\varepsilon \ll 1$ we have Eqs.(\ref{T-gen2}) and (\ref{Q-gen})
without $\varepsilon$-dependence, and
\begin{equation}
    \varepsilon \sim 3.6\times 10^9 \Omega_Q^{-13/5} 
    \beta_s^{4/15} \eta_{B,10} M_{F,6}^{44/15}
    m_{\phi,2}^{2/3}.
\end{equation}

We get the right answer if we equate Eqs.(\ref{B-DM-app}) and
(\ref{Q-gen}), and the required values are as follows:
\begin{equation}
    Q \sim 1.6\times 10^{24}, \quad 
    M_F \sim 4.5\times 10^2 {\rm GeV}, \quad
    \varepsilon \sim 1.0.
\end{equation}
As is the case of the specific potential discussed in the previous
section, the Q-ball charge depends linearly on $\varepsilon$ for 
$\varepsilon \gtrsim 0.1$, and constant for $\varepsilon \lesssim
0.1$. Therefore, we obtain the $M_F$-dependence of Q-ball charge as
\begin{equation}
    Q \propto \left\{
      \begin{array}{ll}
          M_F^4, & M_F \gtrsim 2.5\times 10^2 \ {\rm GeV}, \\[2mm]
          M_F^{12/5}, & M_F \lesssim 2.5\times 10^2 \ {\rm GeV}, \\
      \end{array}
      \right.
\end{equation}
in the range $M_F = 10^2 - 4.5 \times 10^2$ GeV, or, equivalently,
$\varepsilon=6.7\times 10^{-3} - 1.0$ for $m_{\phi}=100$ GeV. Notice
that the condition $T \gtrsim M_F$ must be hold for this scenario to
work properly. It can be rephrased as
\begin{equation}
    M_F \lesssim 6.2 \times 10^4 \Omega_Q^{1/2} \beta_{\ell}^{1/8}
    \ {\rm GeV},
\end{equation}
where Eqs.(\ref{T-gen}) and (\ref{T-M-rela}) are used. Therefore, the
above scenario for $n=6$ is allowed by this condition.

\section{Delayed Q-ball formation}
Since we are interested in the Q-ball formation, we have considered
only $K<0$ cases in the gravity-mediated SUSY breaking effects. 
However, it is possible for some flat directions that $K$ become
positive. As opposed to the gravity-mediation model, the sign of the
$K$ term has an ambiguity because of the complexity of the particle
contents in the gauge-mediated SUSY breaking model. Here, we consider
this possibility and will find some consistent scenario for the dark
matter Q balls providing the baryons of the universe. We will derive
constraints in the generic potential first, and put $M_F=m_{\phi}$ for
the specific one later. 

\subsection{Dominated by the thermal logarithmic potential}
Let us investigate the case which the gravity-mediation and thermal
logarithmic terms dominate the effective potential at large and small
scales, respectively. The critical amplitude of the AD field is
determined by
\begin{equation}
    T^4 \log \frac{\phi^2}{T^2} \sim m_{3/2}^2\phi^2.
\end{equation}
Thus, we get 
\begin{equation}
    \phi_{eq} \sim \frac{T_{eq}^2}{m_{3/2}},
\end{equation}
where the subscript `eq' denotes the variables which are estimated at
the equality of two different potentials above. At this time the
horizon size becomes larger than that at the beginning of the field
rotation. At larger scales when the gravity-mediation term dominates,
the amplitude of the field decreases as $\phi \propto a^{-3/2}$ as the 
universe expands. The universe is assumed to be dominated by
the inflaton-oscillation energy, $a \propto t^{2/3} \propto
H^{-2/3}$, which leads to $\phi \propto H$. We can thus find the
horizon size at $\phi=\phi_{eq}$. It reads as
\begin{equation}
    H_{eq} \sim H_{osc} \frac{\phi_{eq}}{\phi_{ocs}},
\end{equation}
where subscript `osc' denotes the values at the beginning of the field
oscillation (rotation) and $H_{osc} \simeq m_{3/2}$. We find the
horizon size at $\phi=\phi_{eq}$ as
\begin{equation}
    H_{eq}^{-1} \sim \left(\frac{T_{eq}^2}{\phi_0}\right)^{-1}.
\end{equation}
Since the typical size of the resonance mode in the instability band
is $k_{res}^{-1}\sim T_{eq}^2/\phi_{eq}\sim m_{3/2}^{-1}$, the
horizon size is larger, so that the instabilities develop as soon as 
the field feels the negative pressure due to the logarithmic
potential, and Q balls are produced with the typical size 
$\sim k_{res}^{-1}$. Since the angular velocity is written as
$\omega_{eq} \sim (T_{eq}^2/\phi_{eq})^{-1} \sim m_{3/2}$ at this
time, the number density of the field can be expressed as
\begin{equation}
    n_{eq} \sim \omega_{eq}\phi_{eq}^2 \sim \frac{T_{eq}^4}{m_{3/2}}. 
\end{equation}
This corresponds exactly to the number density just diluted by the
cosmic expansion: 
\begin{equation}
    n_{eq} \sim n_{osc}\left(\frac{a_{osc}}{a_{eq}}\right)^3
    \sim n_{osc}\left(\frac{\phi_{osc}}{\phi_{eq}}\right)^{-2}
    \sim \frac{T_{eq}^4}{m_{3/2}}.
\end{equation}

Therefore, the charge of the produced Q ball will be 
\begin{equation}
    Q \sim n_{eq} k_{res}^{-3}
    \sim \left(\frac{T_{eq}}{m_{3/2}}\right)^4
    \sim \left(\frac{\phi_{eq}}{m_{3/2}}\right)^2
    \sim \left(\frac{\phi_{eq}}{T_{eq}}\right)^4,
\end{equation}
where the last and the second last terms have the same forms as the
formulas of the charge estimation for the `usual' and `new' type of
the Q balls, respectively. We thus apply the numerical results which
we obtained for the `usual' type, and assume the charge of the Q ball
as
\begin{equation}
    \label{charge-K}
    Q = \beta \left(\frac{\phi_{eq}}{T_{eq}}\right)^4,
\end{equation}
where $\beta \approx 6 \times 10^{-4}$.
\footnote{%
$\beta$ may be larger than this value, since the cosmic expansion is
weaker than the situations which are done in numerical calculations
above, and the Q-ball formation takes place earlier. However, the
following estimates do not change because of the weak dependence on
$\beta$.}
Notice that 
$\varepsilon \sim 1$ in this case, since the oscillation (rotation) of 
the field starts when the potential is dominated by the
gravity-mediation term where $\omega \sim m_{3/2}$, as mentioned at
the end of Sect.III.

The temperature of the radiation at $\phi=\phi_{eq}$ is determined by
$T_{eq} \sim (M T_{RH}^2 H_{eq})^{1/4}$. We thus find it as
\begin{equation}
    \label{temp-K}
    T_{eq} \sim \sqrt{\frac{M}{\phi_0}}T_{RH}.
\end{equation}

The baryon-to-photon ratio can again be expressed in two ways. One is
\begin{equation}
    \eta_B \sim \frac{\rho_{c,0}\Omega_Q\Delta Q}
                     {n_{\gamma,0}M_F Q^{3/4}},
\end{equation}
and we can rewrite the ratio $r_B$ as
\begin{equation}
    \label{r-B-K}
    r_B \equiv \frac{\Delta Q}{Q} \sim 10^{14} \eta_B \Omega_Q^{-1} 
    \left(\frac{M_F}{10^6 \ {\rm GeV}}\right) Q^{-1/4}.
\end{equation}
On the other hand, we have
\begin{equation}
    \label{eta-B-K}
    \eta_B \sim \frac{r_B n_{\phi,osc}}{\rho_{I,osc}/T_{RH}}
    \sim \frac{r_B\phi_0^2T_{RH}}{m_{3/2}M^2}.
\end{equation}
Inserting Eqs.(\ref{charge-K}), (\ref{temp-K}), and (\ref{r-B-K}) into 
this equation, we obtain the initial amplitude of the field:
\begin{equation}
    \label{phi-0-K}
    \phi_0 \sim 2.9 \times 10^{12} \Omega_Q^{2/5} \beta_{\ell}^{1/10}
    M_{F,6}^{-2/5} \ {\rm GeV}.
\end{equation}
Then the temperature of the radiation and the amplitude of the field
at $\phi=\phi_{eq}$ are 
\begin{eqnarray}
    T_{eq} & \sim & 9.2\times 10^7 \Omega_Q^{-1/5}
    \beta_{\ell}^{-1/20} M_{F,6}^{1/5} T_{RH,5} \ {\rm GeV}, \\
    \phi_{eq} & \sim & 8.4\times 10^{14} \Omega_Q^{-2/5}
    \beta_{\ell}^{-1/10} \nonumber \\
    & & \hspace{15mm} \times M_{F,6}^{2/5} T_{RH,5}^2
    m_{3/2,{\rm GeV}}^{-1} \ {\rm GeV},
\end{eqnarray}
respectively, where $m_{3/2,{\rm GeV}}\equiv m_{3/2}/{\rm GeV}$. We
can thus estimate the charge of the Q ball as
\begin{equation}
    Q \sim 4.3\times 10^{26} \Omega_Q^{4/5}\beta_{\ell}^{4/5}
    m_{F,6}^{4/5} T_{RH,5}^4 m_{3/2,{\rm GeV}}^{-4}.
\end{equation}
Notice that we need the constraint $\phi_0 \gtrsim \phi_{eq}$, which
is rephrased as
\begin{equation}
    \phi_0 \gtrsim 1.5\times 10^{14} T_{RH,5} 
    m_{3/2,{\rm GeV}}^{-1/2} \ {\rm GeV},
\end{equation}
for this situation to take place. 

There are two other conditions to be imposed. One of them is that the 
temperature of the radiation at $\phi=\phi_{eq}$ is larger than $M_F$.
Otherwise, the thermal logarithmic term in the potential is not the 
dominant one, which contradicts the assumption of this subsection. 
This constrains the reheating temperature from below, such as
\begin{equation}
    \label{Trh-K}
    T_{RH} \gtrsim 1.1\times 10^3 \Omega_Q^{1/5}
        \beta_{\ell}^{1/20} M_{F,6}^{4/5} {\rm GeV}.
\end{equation}
The second one is that the gravitino mass should be smaller than 1 GeV,
since we are discussing the gauge-mediated SUSY breaking model. As 
will be seen shortly, this constrains the allowed region of $M_F$ from 
below, when we assume that the initial amplitude of the AD field is 
determined by the balance between the (negative) Hubble mass and
nonrenormalizable terms.  From this assumption, we get
$\phi_0 \sim (m_{3/2}M^{n-3})^{1/(n-2)}$. For some $n$, we can 
write them down as
\begin{equation}
    \label{H-NR-init-phi}
    \phi_0 \sim \left\{
      \begin{array}{ll}
          1.5\times 10^9    m_{3/2,{\rm GeV}}^{1/2} \ {\rm GeV} 
            & (n=4), \\[2mm]
          1.8\times 10^{12} m_{3/2,{\rm GeV}}^{1/3} \ {\rm GeV} 
            & (n=5), \\[2mm]
          6.1\times 10^{13} m_{3/2,{\rm GeV}}^{1/4} \ {\rm GeV} 
            & (n=6), \\[2mm]
          5.1\times 10^{14} m_{3/2,{\rm GeV}}^{1/5} \ {\rm GeV} 
            & (n=7), \\[2mm]
          2.1\times 10^{15} m_{3/2,{\rm GeV}}^{1/6} \ {\rm GeV} 
            & (n=8). \\
      \end{array}
    \right.
\end{equation}
Equating these with Eq.(\ref{phi-0-K}) and using the condition that
the gravitino mass should be less than 1 GeV, we obtain the
constraints on $M_F$. 

Combining these with the condition that the reheating temperature must 
be larger than $\sim 10$ MeV for the sake of the successful
nucleosynthesis \cite{KaKoSu}, we obtain the lower limit of $M_F$ for
the allowed range, using Eq.(\ref{Trh-K}). In general, parameter space
$(Q,M_F)$ is constrained by three conditions: the stability, survival,
and B-DM conditions, which we showed in the last section. See
Eqs.(\ref{stable-app}), (\ref{survive-app}), and (\ref{B-DM-app}). We
find the allowed range of $M_F$ only for $n=6$ and $n=7$, if we take
into account all the conditions which we have mentioned above. These
are
\begin{eqnarray}
    M_F & \sim &4.9\times 10^2 - 1.1\times 10^4 \ {\rm GeV} 
    \quad (n=6), \\ 
    M_F & \sim &1.0\times 10^2 - 2.5\times 10^3 \ {\rm GeV} 
    \quad (n=7). 
\end{eqnarray}
For $n=6$, the lower and upper limits come from the conditions
$m_{3/2}\lesssim 1$ GeV and $M_F < T_{eq}$, respectively. On the other 
hand, for $n=7$, the upper limit comes from the condition 
$T_{RH} \gtrsim 10$ MeV, and we assume that $M_F \gtrsim 100$ GeV.
Therefore, we have consistent scenarios in this model, and, for the
above ranges for $M_F$, the allowed parameter regions are 
\begin{eqnarray}
    Q & \sim & 1.3\times 10^{24} - 2.8\times 10^{21}, \\
    T_{RH} & \sim & 1.0\times 10^7 - 29 \ {\rm GeV}, 
\end{eqnarray}
for $n=6$, and
\begin{eqnarray}
    Q & \sim & 3.2\times 10^{25} - 5.1\times 10^{22}, \\
    T_{RH} & \sim & 62 \ {\rm GeV} - 10 \ {\rm MeV}, 
\end{eqnarray}
for $n=7$, if we use the B-DM condition in order for the dark matter Q 
balls to supply the baryons in the universe. The allowed regions will
be plotted in Figs.~\ref{fig15} and \ref{fig16} in the next section,
where the constraints from several experiments are also plotted.

\subsection{Dominated by zero temperature generic logarithmic term}
Now we will consider the situation when the temperature of the
radiation is rather low, and the effective potential is dominated by
the generic logarithmic term
\begin{equation}
    V \sim M_F^4 \log \left(\frac{\phi^2}{M_S^2}\right).
\end{equation}
Similar  discussion follows along the line which we made in the last
subsection. The critical amplitude of the field is determined by
\begin{equation}
    M_F^4 \log \left(\frac{\phi_{eq}^2}{M_S^2}\right)
    \sim m_{3/2}^2\phi_{eq}^2,
\end{equation}
and we thus get $\phi_{eq}\sim M_F^2/m_{3/2}$. The horizon size at
$\phi=\phi_{eq}$ is then written as 
\begin{equation}
    H_{eq} \sim H_{osc}\frac{\phi_{eq}}{\phi_0} 
    \sim \frac{M_F^2}{\phi_0},
\end{equation}
which is larger than the typical resonance scale in the instability
band: $k_{res}^{-1}\sim (M_F^2/\phi_{eq})^{-1}$. Therefore, the field
feels spatial instabilities just after its amplitude gets smaller than
$\phi_{eq}$. The charge of the Q ball can be estimated as
\begin{equation}
    \label{Q-K-2}
    Q = \beta \frac{\phi_{eq}^4}{M_F^4} 
    \sim 6.0\times 10^{20} 
      \beta_{\ell} M_{F,6}^4 m_{3/2,{\rm GeV}}^{-4}.
\end{equation}

Using Eqs.(\ref{r-B-K}), (\ref{eta-B-K}), and (\ref{Q-K-2}), we obtain 
the initial amplitude required for the Q-ball formation in this
scenario as
\begin{equation}
    \label{phi-0-MF}
    \phi_0 \sim 9.5\times 10^{10} \Omega_Q^{1/2}\beta_{\ell}^{1/8}
    T_{RH,6}^{-1/2} \ {\rm GeV},
\end{equation}
where $T_{RH,6}\equiv T_{RH}/(10^6 {\rm GeV})$. 

At the time of the production of Q balls when $\phi \sim \phi_{eq}$,
the temperature of the radiation can be estimated from
$T_{eq} \sim (MT_{RH}H_{eq})^{1/4}$, so that we have
\begin{equation}
    T_{eq} \sim 7.1\times 10^7 \Omega_Q^{-1/8}\beta_{\ell}^{-1/32}
    T_{RH,6}^{5/8}M_{F,6}^{1/2} \ {\rm GeV}.
\end{equation}
The condition that the initial amplitude of the field exceeds the
critical value where the effects of gauge- and gravity-mediation on
the effective potential are comparable is expressed as
\begin{equation}
    \phi_0 \gtrsim 10^{12} M_{F,6}^2 m_{3/2,{\rm GeV}}^{-1}
    \ {\rm GeV}.
\end{equation}

Another condition for this situation to be realized is 
$M_F\gtrsim T_{eq}$, which can be rewritten as 
\begin{equation}
    \label{MF-a}
    M_F \gtrsim 5.0\times 10^9 \Omega_Q^{-1/4}\beta_{\ell}^{-1/16}
    T_{RH,6}^{5/4} \ {\rm GeV}.
\end{equation}
This implies that the allowed region in the parameter space $(Q,M_F)$
can appear for low enough reheating temperature. It can be seen as
follows. From the stability and B-DM conditions
[Eqs.(\ref{stable-app}) and (\ref{B-DM-app})], the largest possible
$M_F$ is 
\begin{equation}
    M_F \sim 8.2\times 10^4 \varepsilon^{1/4} \Omega_Q^{1/4}
    \eta_{B,10}^{-1/4}m_{\phi,2}^{-1/6} \ {\rm GeV}.
\end{equation}
Therefore, comparing these two, we get the constraint on the reheating
temperature as
\begin{equation}
    T_{RH} \lesssim 1.5\times 10^2  \varepsilon^{1/5} \Omega_Q^{2/5}
    \beta_{\ell}^{1/20} \eta_{B,10}^{-1/5} m_{\phi,2}^{-2/15} 
    \ {\rm GeV},
\end{equation}
for generating the allowed region. On the other hand, we can also
obtain the lower limit of $M_F$. In order to have successful
nucleosynthesis, we need $T_{RH}\gtrsim 10$ MeV conservatively. We
thus get the lower bound as $M_F \gtrsim 0.5$ GeV from
Eq.(\ref{MF-a}). However, we will assume $M_F \gtrsim 100$ GeV, which
may be the lowest possible scale for SUSY breaking, and this limit is
severer. 

At the same time, we can estimate the largest possible gravitino mass,
taking into account the fact that the charge of the produced Q ball
[Eq.(\ref{Q-K-2})] should satisfy the stability condition
[Eq.(\ref{stable-app})], as 
\begin{equation}
    m_{3/2} \lesssim 0.16 \beta_{\ell}^{1/4} \ {\rm GeV}. 
\end{equation}
Therefore, we obtain the allowed values of parameters as
\begin{eqnarray}
    \label{allowed-condition-1}
    & & m_{3/2} \lesssim  0.16  \ {\rm GeV}, \\
    \label{allowed-condition-2}
    & & 10 \ {\rm MeV} \lesssim T_{RH} \lesssim 110 \ {\rm GeV}, \\
    \label{allowed-condition-3}
    & & M_F \gtrsim 100 \ {\rm GeV}, 
\end{eqnarray}
in general.

The initial amplitude of the field is the last to be considered. The
initial value is determined by the balance of the Hubble mass and
nonrenormalizable terms, so we have exactly same formulas as in 
Eq.(\ref{H-NR-init-phi}). It depends only on the gravitino mass as
$\phi_0 \sim (m_{3/2}M^{n-3})^{1/(n-2)}$. Equating with
Eq.(\ref{phi-0-MF}), we get the relation between the gravitino mass
and the reheating temperature. As easily seen, there is no allowed
range for $n=4$ and 5, while we have
\begin{eqnarray}
    490 \ {\rm keV} \lesssim & m_{3/2} & \lesssim 0.16 \ {\rm GeV}, 
    \nonumber \\
    1.1\times 10^2 \ {\rm GeV} \gtrsim & T_{RH} 
                                    & \gtrsim 6.1 \ {\rm GeV},  
\end{eqnarray}
for $n=6$,
\begin{eqnarray}
    1.8  \ {\rm eV} \lesssim & m_{3/2} & \lesssim 0.16 \ {\rm GeV}, 
    \nonumber \\
    1.1\times 10^2 \ {\rm GeV} \gtrsim & T_{RH} 
                                    & \gtrsim 72 \ {\rm MeV},  
\end{eqnarray}
for $n=7$,
\begin{eqnarray}
    6.4\times 10^{-15}  \ {\rm GeV} \lesssim & m_{3/2} 
            & \lesssim 8.6\times 10^{-3} \ {\rm GeV}, \nonumber \\
    1.1\times 10^2 \ {\rm GeV} \gtrsim & T_{RH} 
                                    & \gtrsim 10 \ {\rm MeV},  
\end{eqnarray}
for $n=8$, and so on. 

On the other hand, we can get the value of $M_F$ required for both the 
Q-ball formation [Eq.(\ref{Q-K-2})] and the B-DM condition, in terms
of the gravitino mass as 
\begin{equation}
    M_{F,6} \sim 0.28 \varepsilon^{1/4}\Omega_Q^{1/4}
    \beta_{\ell}^{-1/6}\eta_{B,10}^{-1/4}m_{\phi,2}^{-1/6}
    m_{3/2,{\rm GeV}}^{2/3}.
\end{equation}
Inserting the $m_{3/2}-T_{RH}$ relation for each $n$ into this
equation, we have 
\begin{equation}
    M_F \sim \left\{ 
      \begin{array}{ll}
          9.2\times 10^{-3} \Omega_Q^{19/12}m_{\phi,2}^{-1/6} 
            T_{RH,6}^{-4/3} \ {\rm GeV}, & (n=6), \\[2mm]
          1.1\times 10^{-7} \Omega_Q^{23/12}m_{\phi,2}^{-1/6} 
            T_{RH,6}^{-5/3} \ {\rm GeV}, & (n=7), \\[2mm]
          1.2\times 10^{-12} \Omega_Q^{9/4}m_{\phi,2}^{-1/6} 
            T_{RH,6}^{-2} \ {\rm GeV}, & (n=8), \\[2mm]
          1.4\times 10^{-17} \Omega_Q^{31/12}m_{\phi,2}^{-1/6} 
            T_{RH,6}^{-7/3} \ {\rm GeV}, & (n=9), \\
      \end{array}
      \right.
\end{equation}
where we suppress the dependences on $\beta$, $\eta_B$, and
$\varepsilon$, since they are trivial. These values have to satisfy
the constraint Eq.(\ref{MF-a}). They lead to the upper limit for the
reheating temperature, and we obtain allowed regions only for $n=6$,
7, and 8, taking into account constraints (\ref{allowed-condition-1}),
(\ref{allowed-condition-2}), and (\ref{allowed-condition-3}). Thus, we 
find the allowed region for the consistent scenarios as
\begin{eqnarray}
    6.1 \ {\rm GeV} \lesssim & T_{RH} & \lesssim 29 \ {\rm GeV}, 
      \nonumber \\
    8.3\times 10^4 \ {\rm GeV} \gtrsim & M_F 
      & \gtrsim 1.0\times 10^4 \ {\rm GeV},\nonumber \\
    0.16 \ {\rm GeV} \gtrsim & m_{3/2} 
      & \gtrsim 7.0\times 10^{-3} \ {\rm GeV},
\end{eqnarray}
for $n=6$,
\begin{eqnarray}
    72 \ {\rm MeV} \lesssim & T_{RH} & \lesssim 1.9 \ {\rm GeV}, 
      \nonumber \\
    8.3\times 10^4 \ {\rm GeV} \gtrsim & M_F 
      & \gtrsim 3.6\times 10^2 \ {\rm GeV},\nonumber \\
    0.16 \ {\rm GeV} \gtrsim & m_{3/2} 
      & \gtrsim 4.5\times 10^{-5} \ {\rm GeV},
\end{eqnarray}
for $n=7$, and
\begin{eqnarray}
    10 \ {\rm MeV} \lesssim & T_{RH} & \lesssim 110 \ {\rm MeV}, 
      \nonumber \\
    1.2\times 10^4 \ {\rm GeV} \gtrsim & M_F 
      & \gtrsim 10^2 \ {\rm GeV},\nonumber \\
    8.6 \ {\rm MeV} \gtrsim & m_{3/2} 
      & \gtrsim 6.4 \ {\rm keV},
\end{eqnarray}
for $n=8$, when we take $m_{\phi}=100$ GeV.

\section{Detection of the dark matter Q-ball}
\subsection{Q balls in the specific logarithmic potential}
As is well known, Q balls are stable against the decay into nucleons
in the gauge mediation mechanism for SUSY breaking
\cite{KuSh,KK3}. Therefore, they can be considered as the good
candidate for the dark matter of the universe. In addition, they
can supply the baryon number of the universe. In the previous
sections, we investigate if there is any consistent cosmological
scenario for the dark matter Q ball with simultaneously supplying the
baryon number of the universe. We have also seen the amount of the
charge evaporated from a Q ball, so we can relate it to the amount of
the dark matter. In this section, we see more conservative allowed
region for the cosmological Q-ball scenario to work, and impose the
experimental bounds in order to see if the scenario could exist.

Provided that the initial charge of the Q ball is larger than the
evaporated charge, we regard that the Q ball survives from
evaporation, and contributes to the dark matter of the universe. This
is expressed in Eq.(\ref{survive-2})

\begin{equation}
    \label{survive}
    Q_{{\rm init}} \gtrsim 7.4\times 10^{17} 
           \left( \frac{m_{\phi}}{{\rm TeV}} \right)^{-12/11}.
\end{equation}

Now we can relate the baryon number and the amount of the dark matter
in the universe. As mentioned above, the baryon number of the universe
should be explained by the amount of the charge evaporated from the Q
balls, $\Delta Q$, and the survived Q balls become the dark matter. 
This condition is Eq.(\ref{dm-Q}):
\begin{equation}
    \label{dm}
    Q \lesssim 10^{23} \varepsilon^{3/2}
              \left( \frac{m_{\phi}}{{\rm TeV}} \right)^{-3},
\end{equation}
where we take $\eta_B \sim 10^{-10}$ and $\Omega_Q \lesssim 1$.

In order for the Q ball to be stable against the decay into nucleons, 
i.e., $E_Q/Q\lesssim 1$ GeV, the following should be satisfied: 
\begin{equation}
    \label{stable}
    Q \gtrsim 10^{12} \left( \frac{m_{\phi}}{{\rm TeV}} \right)^4.
\end{equation}

\begin{figure}[t!]
\centering
\hspace*{-7mm}
\leavevmode\epsfysize=12cm \epsfbox{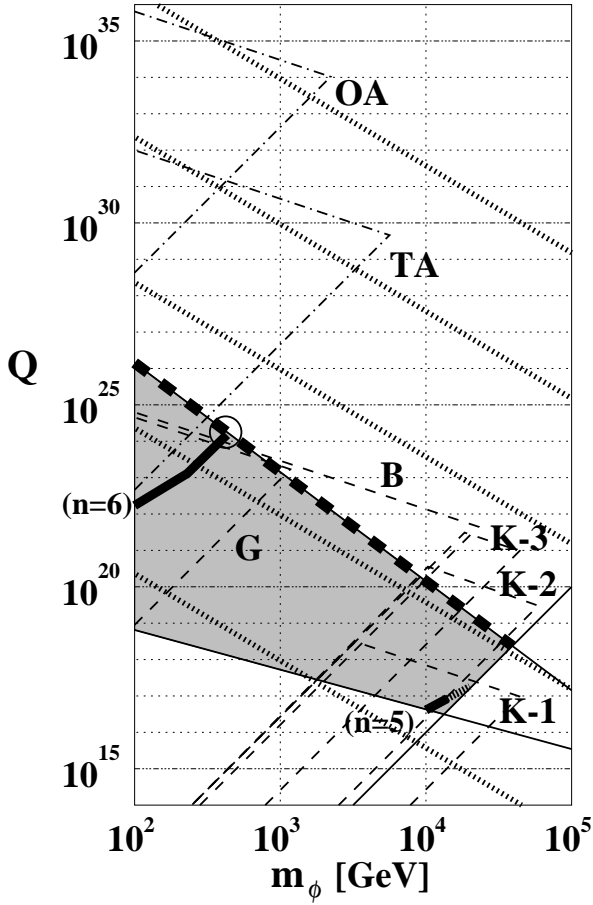}\\[2mm]
\caption[fig12]{\label{fig12} 
Summary of constraints on ($Q,m_{\phi}$) plain for the {\it usual}
type of the Q ball. We also show the regions currently excluded by
BAKSAN (B), Gyrlyand (G), and Kamiokande (K-1, K-2, K-3), and to be
searched by the Telescope Array Project (TA) and OWL-AIRWATCH (OA) in
the future. For the details of experiments, see Ref.~\cite{ArYoNaOg}.
Two thick lines within the shaded region denote the consistent
scenarios for $n=5$ and $n=6$ cases, respectively, and the only
allowed consistent scenario is found inside the circle. Five thick
dotted lines denote the constraints of the dark matter Q ball which do
not account for the baryogenesis for $T_{RH}=10^2$, $10^3$, $10^4$,
$10^5$, and $10^6$ GeV from the top to the bottom.}
\end{figure}

For the {\it usual} gauge-mediation type of the Q ball, we obtain the
allowed region for explaining the baryon number of the universe, using 
three constraints, i.e., Eqs. (\ref{survive}),(\ref{dm}), and
(\ref{stable}). Figure \ref{fig12} shows the allowed region on
$(Q,m_{\phi})$ plane with $\varepsilon=1$ in Eq.(\ref{dm}) \cite{KK5}. 
The shaded regions represent that this type of stable Q balls are
created, and the baryon number of the universe can be explained by the
mechanism mentioned above. Furthermore, this type of stable Q balls
contribute crucially to the dark matter of the universe at present, if
the Q balls have the charge given by the thick dashed line in the
figure. Notice that this line denotes $\varepsilon=1$ case, and the
allowed region in the parameter space will be narrower as
$\varepsilon$ becomes smaller, and disappear for 
$\varepsilon \lesssim 10^{-5}$.

As can be seen in Fig.~\ref{fig12}, several experiments constrain the
parameter space. The Q ball can be detected through so-called 
Kusenko-Kuzmin-Shaposhnikov-Tinyakov (KKST) process. When nucleons
collide with a Q ball, they enter the surface layer of the Q ball,
and dissociate into quarks, which are converted into squarks. In this
process, Q balls release $\sim 1$ GeV energy per collision by emitting
soft pions. This is the basis for the Q-ball detections
\cite{KuKuShTi,ArYoNaOg}. Lower left regions are excluded by the
various experiments. The allowed charges are $Q \sim 10^{25}$ with
$m_{\phi}=100$ GeV $- 1$ TeV for $\varepsilon=1$, and
future experiments such as the Telescope Array Project or the
OWL-AIRWATCH detector may detect the dark matter Q balls. 

We also put the consistent scenario in Fig.~\ref{fig12}, which we
considered in Sec.~\ref{scenario}. Two thick lines inside the allowed 
region (shaded region) represent for the $n=5$ and $n=6$ cases. Hatched
line connected to the thick line for $n=5$ shows the allowed region if
we do not assume that the A terms come from the nonrenormalizable
superpotential. The current experiments exclude the $n=5$ case, but the
top edge of $n=6$ case is allowed, which we show by circle in the
figure. Although the $n=6$ case sits at very interesting region, this
dark matter Q ball may not be detected by future experiments as
mentioned above. 

If we do not impose the condition that the evaporated charges from Q
balls account for the baryons in the universe, the only constraint is
that the energy density of Q balls must not exceed the critical
density. As mentioned in the previous section, this condition is 
Eq.(\ref{Q-limit}):
\begin{eqnarray}
    Q \sim 9.3 \times 10^{21} & & \varepsilon^{8/5} \Omega_Q^{12/5}
    \left(\frac{T_{RH}}{10^5 {\rm GeV}}\right)^{-4} \nonumber \\
    & &  \times
    \left(\frac{\beta}{6\times 10^{-4}}\right)^{8/5}
    \left(\frac{m_{\phi}}{\rm TeV}\right)^{-12/5}.
\end{eqnarray}
The Q-ball charge depends on the reheating temperature, and thick
dotted parallel lines show the constraints for $T_{RH}=10^2$, $10^3$,
$10^4$, $10^5$, and $10^6$ GeV from the top to the bottom in the
figure. We can thus discover the dark matter Q balls in the future
experiments, if the reheating temperature is low enough. Notice that
the lower reheating temperature is favored by the supersymmetric
models because of evading the cosmological gravitino problem
\cite{Moroi}.

\subsection{Q balls in the generic logarithmic potential}

\begin{figure}[t!]
\centering
\hspace*{-7mm}
\leavevmode\epsfysize=11cm \epsfbox{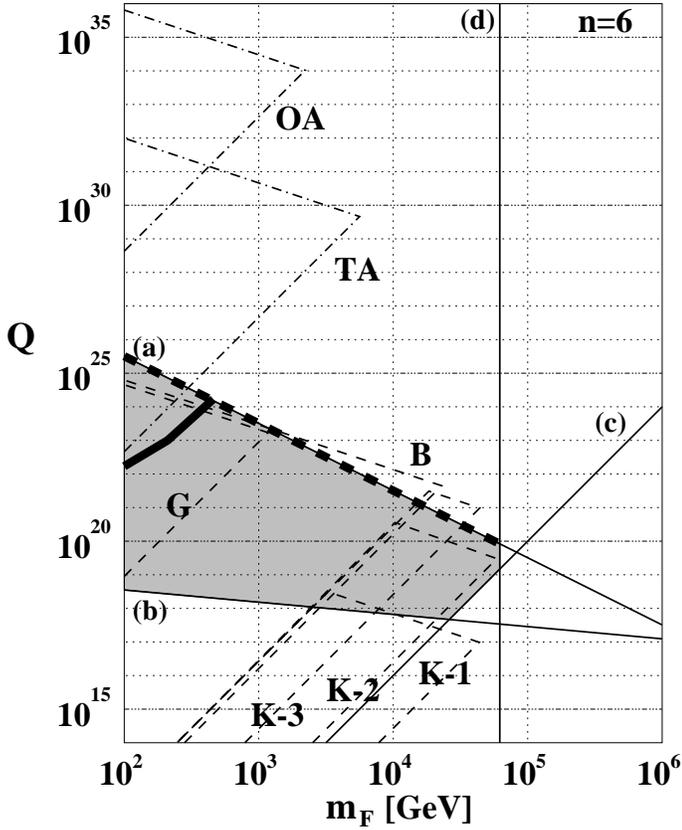}\\[2mm]
\caption[fig13]{\label{fig13} 
Summary of constraints on the parameter space ($Q,M_F$) with $n=6$ and 
$m_{\phi}=100$ GeV in the generic logarithmic potential where the
thermal terms are dominated when the Q-ball formation occurs. Dashed
and dot-dashed lines represent the same current and future
experiments, respectively, as shown in Fig.~\ref{fig12}. Lines (a),
(b), and (c) denote the B-DM, survival, and stability conditions,
respectively. Line (d) shows the conditions $T \gtrsim M_F$. Notice
that this condition depends on $n$. The thick solid line is the
allowed region of the successful scenario for the dark matter Q balls
supplying the baryons in the universe, while the thick dashed line
represents more general cases.} 
\end{figure}

Now we show the allowed region for the Q-ball formation in the generic 
potential, but actual formation takes place when the thermal
logarithmic term still dominates the potential. We show for the $n=6$
case, which is the only successful scenario in this situation, in
Fig.~\ref{fig13}. The thick solid line denotes the natural consistent
scenario, while the thick dashed line represents the allowed region on
the B-DM condition line in a more general situation that the Q-ball
formation occurs through the different mechanism. On these lines, the
Q balls can account for both the dark matter and the baryons in the
universe simultaneously. Most of the parameter space is excluded by
currents experiments, but very interesting region such as 
$Q\sim 10^{24}$ and $M_F\sim 5\times 10^2$ GeV is allowed. Notice that
the dark matter Q balls with larger charges can be detected by the
future experiments such as TA and OA, although they cannot play the
role for the baryogenesis. 

\subsection{Q balls in the gravity-mediation dominated potential}
\begin{figure}[t!]
\centering
\hspace*{-7mm}
\leavevmode\epsfysize=11cm \epsfbox{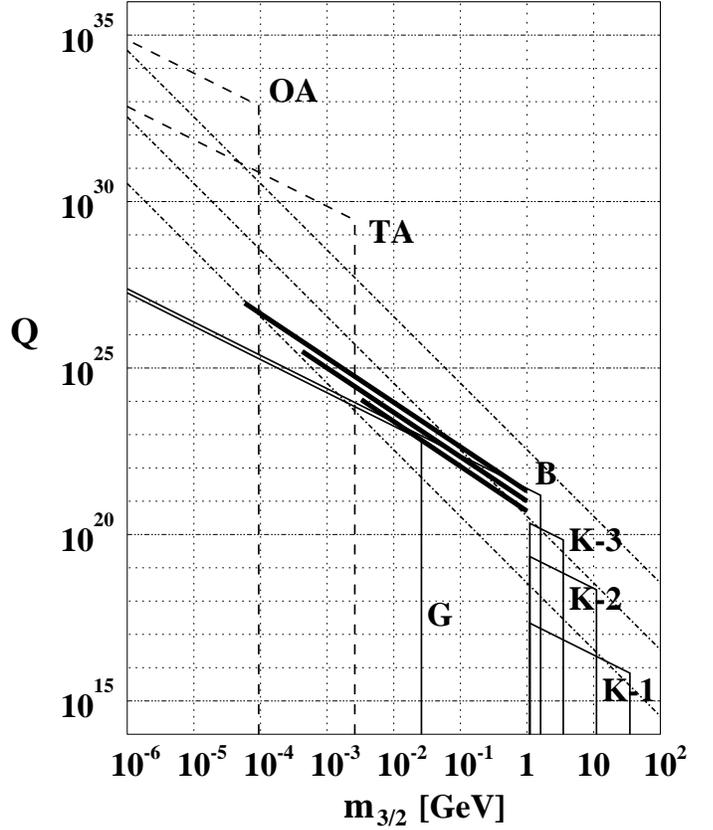}\\[2mm]
%\vspace{-3mm}
\caption[fig14]{\label{fig14} 
Restrictions by several experiments on ($Q,m_{3/2}$) plain
for $|K|=0.01$. We show the regions currently excluded by BAKSAN (B),
Gyrlyand (G), and Kamiokande (K-1, K-2, K-3), and to be searched by
the Telescope Array Project (TA) and OWL-AIRWATCH (OA) in the future. 
The thick lines represent for the gravity-mediation type of the Q ball
to be both the dark matter and the source for the baryons of the
universe for $m_{\phi}=300$ GeV, 1 TeV, and 3 TeV from the top to the
bottom. Three dot-dashed lines denote the constraints of the dark
matter Q ball which do not account for the baryogenesis for 
$T_{RH}=10^3$, $10^5$, and $10^7$ GeV from the top to the bottom.}
\end{figure}

Next, we move on to the `new' type of Q ball. The constraints on
parameter space $(Q,m_{3/2})$ were obtained in Ref.~\cite{KK3}. Here
we only draw the results. The regions where both the amount of the
baryon and the dark matter can be explained are shown as thick lines
in Fig.~\ref{fig14}. Lower left regions are excluded by the various
experiments. Notice that future experiments such as the Telescope
Array Project or the OWL-AIRWATCH detector may detect this type of the 
dark matter Q balls, supplying the baryons in the universe,  with an
interesting gravitino mass $\sim 100$ keV, if the the origin of the A
terms differs from the nonrenormalizable superpotential. 

Although any consistent cosmological Q-ball scenario does not exist in
the framework that both the initial amplitude and the A terms of the
AD field are determined by the nonrenormalizable superpotential, it is
not necessarily true that the dark matter Q balls do not exist whether
they can be the source for the baryons in the universe or not. In
general, we have the constraint for the Q balls to be a crucial
component of the dark matter. It reads as
\begin{equation}
    Q \sim 3.5\times 10^{26} \Omega_Q
    \left(\frac{\beta}{6\times 10^{-4}}\right)
     \left(\frac{m_{3/2}}{\rm MeV}\right)^{-2}
     \left(\frac{T_{RH}}{10^5 {\rm GeV}}\right)^{-1},
\end{equation}
where Eq.(\ref{DM-detect}) is used. 

Three dot-dashed lines are shown for $T_{RH}=10^3$, $10^5$, and
$10^7$, from the top to the bottom, respectively. Therefore, the dark
matter Q balls can be detected in the future experiments for the low
reheating temperature universe, which is free from the cosmological
gravitino problem. Notice that the baryons have to be supplied by 
another mechanism in this case.

\subsection{Delayed Q balls in the thermal logarithmic potential}
\begin{figure}[t!]
\centering
\hspace*{-7mm}
\leavevmode\epsfysize=10cm \epsfbox{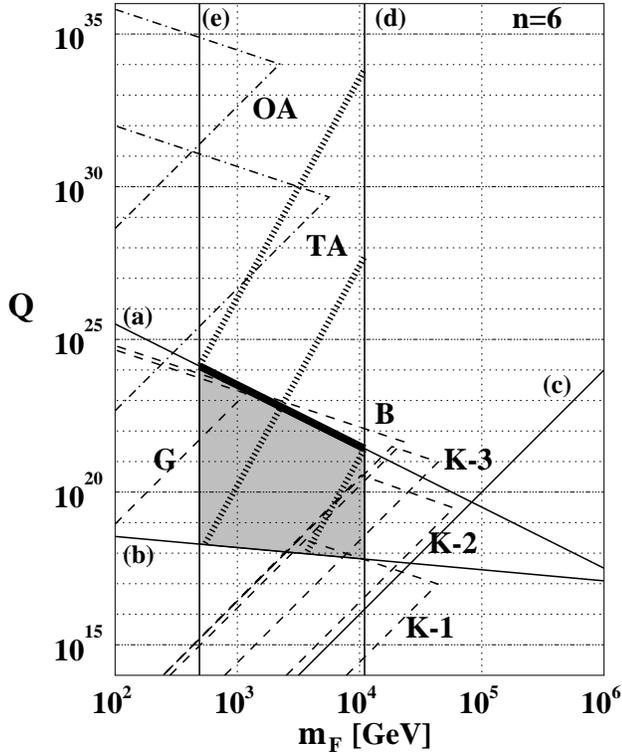}\\[2mm]
%\vspace{-3mm}
\caption[fig15]{\label{fig15} 
Summary of constraints on the parameter space $(Q,M_F)$ for the
delayed-formed Q balls with the thermal logarithmic term domination
for $m_{\phi}=100$ GeV and $n=6$. Dashed and dot-dashed lines
represent the same current and future experiments, respectively, as
shown in Fig.~\ref{fig12}. Lines (a), (b), and (c) denote the B-DM,
survival, and stability conditions, respectively. Lines (d) and (e)
show the conditions $T_{eq} \gtrsim M_F$ and $m_{3/2} \lesssim 1$ GeV,
respectively. The thick solid line is the allowed region of the
successful scenario for the dark matter Q balls supplying the baryons
in the universe. Thick dotted lines denote the formed Q-ball charges
with $T_{RH} \sim 10^7$, $10^3$, and $29$ GeV from the top to the
bottom.
}
\end{figure}

\begin{figure}[t!]
\centering
\hspace*{-7mm}
\leavevmode\epsfysize=10cm \epsfbox{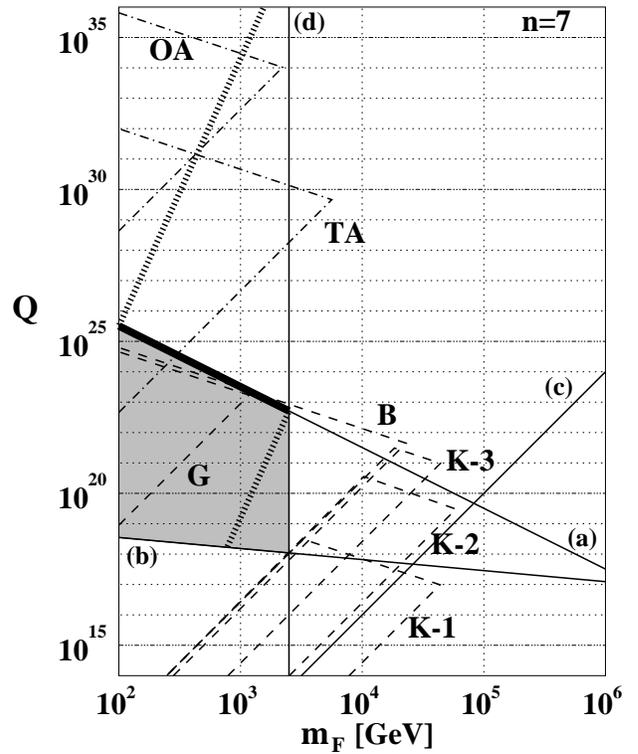}\\[2mm]
%\vspace{-3mm}
\caption[fig16]{\label{fig16} 
Summary of constraints on the parameter space $(Q,M_F)$ for the
delayed-formed Q balls with the thermal logarithmic term domination
for $m_{\phi}=100$ GeV and $n=7$. Dashed and dot-dashed lines
represent the same current and future experiments, respectively, as
shown in Fig.~\ref{fig12}. Lines (a), (b), and (c) denote the B-DM,
survival, and stability conditions, respectively. Line (d) shows the
condition $T_{RH} \gtrsim 10$ MeV. The thick solid line is the allowed
region of the successful scenario for the dark matter Q balls
supplying the baryons in the universe. Thick dotted lines denote the
formed Q-ball charges with $T_{RH} \sim 62$ GeV and $10$ MeV for the
upper and lower lines, respectively. 
}
\end{figure}

As shown in the previous sections, we can restrict the parameter
space $(Q,M_F)$ by several conditions for the Q-ball formation in the
natural scenarios. In general, we have three conditions: the B-DM,
survival, and stability conditions. Here, we write them down again:
\begin{eqnarray}
    {\rm (a)} \quad Q & \sim & 3.2\times 10^{17} \varepsilon^{3/2}
    \Omega_Q^{3/2} \eta_{B,10}^{-3/2} m_{\phi,2}^{-1}
    M_{F,6}^{-2}, \\
    {\rm (b)} \quad Q & \gtrsim & 
    1.2\times 10^{17} m_{\phi,2}^{-8/11} M_{F,6}^{-4/11}, \\
    {\rm (c)} \quad Q & \gtrsim & 10^{24} M_{F,6}^4.
\end{eqnarray}
We plot these lines in Figs.~\ref{fig15} and \ref{fig16} for $n=6$ and
$n=7$ cases, respectively. In these figures, (a), (b), and (c) denote
the B-DM, survival, and stability conditions, respectively, for
$\Omega_Q=1$ and $m_{\phi}=100$ GeV. We also show the data from
several experiments. These are the same as we have used above. The
experimental lines are the same as that for the specific logarithmic
potential, if we just replace $m_{\phi}$ by $M_F$. We plot these data,
using the fact that the flux is related to the Q-ball mass, and the
cross section is related to the Q-ball size. Therefore, these Q-ball
parameters can be expressed in the same form as in the specific
potential for the replacement of $m_{\phi}$ by $M_F$ in the generic 
potential. Notice that the presence of the dark matter Q balls for
$m_{\phi}=1$ TeV is almost excluded by experiments: Only Q balls with
$Q\simeq 10^{19}$ and $M_F \simeq 5.6\times 10^4$ GeV are allowed.

Now we put the conditions for the Q-ball formation to take place
in the natural scenarios. In Fig.~\ref{fig15}, vertical line (d)
denotes for the condition that the thermal logarithmic potential
dominates over the generic ones, i.e., $T_{eq} \gtrsim M_F$. Another
vertical line (e) represents for $m_{3/2} \lesssim 1$ GeV in the
gauge-mediated SUSY breaking model. Three thick dotted lines show the
charges of the produced Q balls in the natural scenarios with the
reheating temperature $1.0\times 10^7$, $1.0\times 10^3$, and 
$29$ GeV from the top to the bottom, respectively. The thick solid
line is the final answer, which represents the dark matter Q balls
supplying the baryons in the universe in the natural scenario.
Two remarks are following. (1) Those dark matter Q balls which do
not account for the baryogenesis may be detected by the Telescope
Array Project (TA), if their charges are $\sim 10^{29}$ and the
reheating temperature is $\sim 10^7$ GeV. (2) There is no cosmological
gravitino problem for this scenario.

In Fig.~\ref{fig16}, line (d) denotes the condition that the reheating 
temperature should be higher than $10$ MeV in order for the
nucleosynthesis to take place successfully. We also assume 
$M_F \gtrsim 100$ GeV. Two thick dotted lines represent the Q-ball
charges from the formation mechanism with the reheating temperature 
62 GeV and 10 MeV for upper and lower lines, respectively. Thick solid 
line shows the successful region that the Q balls account for both
the dark matter and the source for the baryons in the universe
simultaneously. Most of this region is allowed experimentally, and, in 
particular, the dark matter Q balls with $Q \sim10^{25}$ may be
detected by the TA in the future. Notice again that cosmological
gravitino problem is avoided for such low reheating temperature. 

\begin{figure}[t!]
\centering
\hspace*{-7mm}
\leavevmode\epsfysize=10cm \epsfbox{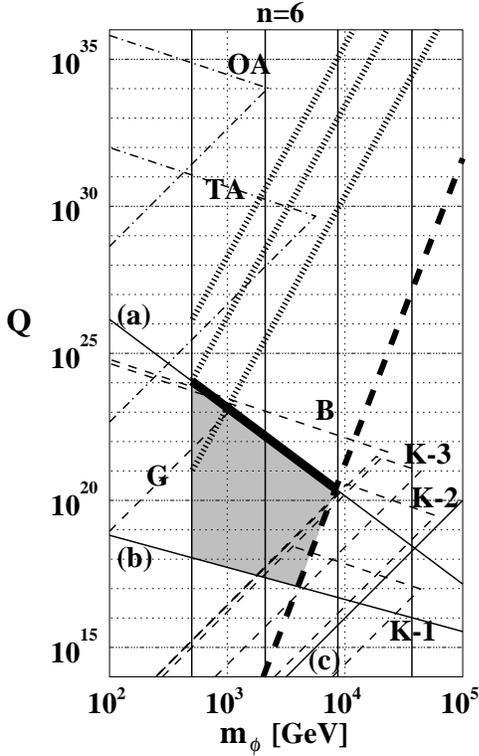}\\[2mm]
%\vspace{-3mm}
\caption[fig17]{\label{fig17} 
Summary of constraints on the parameter space $(Q,m_{\phi})$ for the
delayed-formed Q balls with the thermal logarithmic term domination
for $n=6$. Experimental lines are the same. Lines (a), (b), and (c)
denote the B-DM, survival, and stability conditions, respectively. 
Thick dotted lines denote the formed Q-ball charges for 
$T_{RH} \simeq 10^5$, $3\times 10^4$, and $5\times 10^3$ GeV from the
top to the bottom. The thick dashed line represents the condition 
$T_{eq} \gtrsim m_{\phi}$. The thick solid line is the allowed region
of the successful scenario for the dark matter Q balls supplying the
baryons in the universe. Solid vertical lines denote the gravitino
mass ranges, $m_{3/2}=1$ GeV, 100 MeV, 10 MeV, and 1 MeV from the left
to the right.
}
\end{figure}

\begin{figure}[t!]
\centering
\hspace*{-7mm}
\leavevmode\epsfysize=10cm \epsfbox{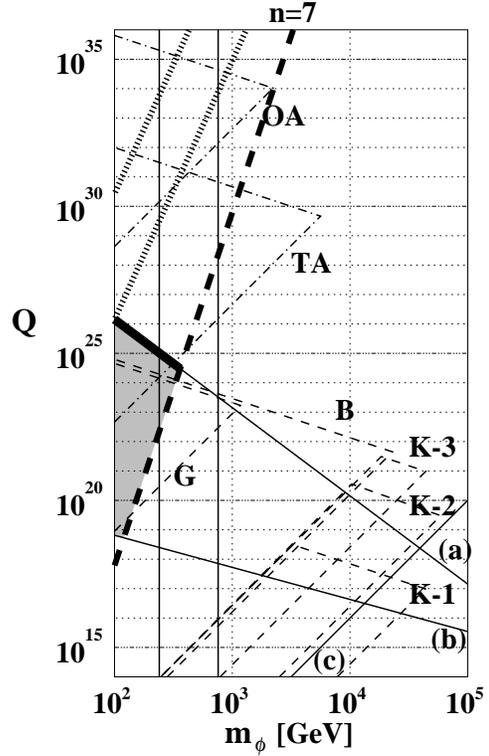}\\[2mm]
%\vspace{-3mm}
\caption[fig18]{\label{fig18} 
Summary of constraints on the parameter space $(Q,m_{\phi})$ for the
delayed-formed Q balls with the thermal logarithmic term domination
for $n=7$. Experimental lines are the same. Lines (a), (b), and (c)
denote the B-DM, survival, and stability conditions, respectively. 
Thick dotted lines denote the formed Q-ball charges for 
$T_{RH} \simeq 10^3$ and $80$ GeV for the upper and lower lines. The
thick dashed line represents the condition $T_{eq} \gtrsim m_{\phi}$. 
The thick solid line is the allowed region of the successful scenario
for the dark matter Q balls supplying the baryons in the
universe. Solid vertical lines denote the gravitino mass ranges,
$m_{3/2}=100$ keV (left) and 10 keV (right).
}
\end{figure}

Now we will mention the specific logarithmic potential, where
$M_F=m_{\phi}$. We impose three general conditions; namely, the
B-DM, survival, and stability conditions. They are written as [cf. 
Eqs.(\ref{dm}), (\ref{survive}), and (\ref{stable})]
\begin{eqnarray}
    {\rm (a)} \quad Q & \sim & 
       1.2\times 10^{23} \varepsilon^{3/2} \Omega_Q^{3/2}
       \eta_{B,10}^{-3/2} m_{\phi,{\rm TeV}}^{-3}, \\
    {\rm (b)} \quad Q & \gtrsim & 
       7.4\times 10^{17} m_{\phi,{\rm TeV}}^{-12/11}, \\
    {\rm (c)} \quad Q & \gtrsim & 10^{12} m_{\phi,{\rm TeV}}^4,
\end{eqnarray}
which are shown in Figs.~\ref{fig17} and \ref{fig18}.

The charge of the formed Q-ball and the initial amplitude of the field
are expressed as 
\begin{eqnarray}
    \label{Q-form-pK}
    Q & \sim & 1.7\times 10^{26} \Omega_Q^{4/5} \beta_{\ell}^{4/5}
    m_{\phi,{\rm TeV}}^{4/5} T_{RH,5}^4 m_{3/2,{\rm GeV}}^{-4}, \\
    \label{phi-0-form-pK}
    \phi_0 & \sim & 5.3\times 10^{14} \Omega_Q^{-2/5} 
    \beta_{\ell}^{-1/10} \nonumber \\ & & \hspace*{15mm}\times
    m_{\phi,{\rm TeV}}^{2/5} T_{RH,5}^2
    m_{3/2,{\rm GeV}}^{-1} {\rm GeV},
\end{eqnarray}
respectively. The initial amplitude is determined by the balance
between the Hubble mass and nonrenormalizable terms, and combining it
with Eq.(\ref{phi-0-form-pK}), we get the relation between $m_{\phi}$
and $m_{3/2}$ as
\begin{equation}
    m_{\phi} \sim 4.9\times 10^2 \Omega_Q \beta_{\ell}^{1/4}
    m_{3/2,{\rm GeV}}^{-5/8} \ {\rm GeV},
\end{equation}
for $n=6$. We thus get the charge of the produced Q balls:
\begin{equation}
    Q \sim 1.9\times 10^{28} \Omega_Q^{-28/5}\beta_{\ell}^{-4/5}
    T_{RH,5}^4 m_{\phi,{\rm TeV}}^{36/5},
\end{equation}
which we plot for $T_{RH} = 10^5$, $3\times 10^4$, and $5\times 10^3$
GeV from the top to the bottom by thick dotted lines in
Fig.~\ref{fig17}. In order for this situation to take place, 
$T_{eq} \gtrsim m_{\phi}$, which leads to the constraint for the
reheating temperature in terms of $m_{\phi}$ as 
\begin{equation}
    \label{Trh-m-phi}
    T_{RH} \gtrsim 4.4 \Omega_Q^{1/5} \beta_{\ell}^{1/20}
    m_{\phi,{\rm TeV}}^{4/5} \ {\rm GeV}.
\end{equation}
Inserting this constraint into Eq.(\ref{Q-form-pK}), we have
\begin{equation}
    Q \gtrsim 6.8\times 10^{10} \Omega_Q^{-24/5}\beta_{\ell}^{-3/5}
    m_{\phi,{\rm TeV}}^{52/5},
\end{equation}
which is shown by the thick dashed line in Fig.~\ref{fig17}. As is
seen in the figure, the experimentally allowed region is very narrow,
and the consistent natural scenario will work only for 
$Q\sim 10^{24}$, $m_{\phi} \sim 5\times 10^2$ GeV, $m_{3/2}\sim 1$
GeV, and $T_{RH} \sim 3\times 10^4$ GeV. Notice that there is no
cosmological gravitino problem also in this case.

On the other hand, the relation between $m_{\phi}$ and $m_{3/2}$,
which is obtained by equating Eq.(\ref{phi-0-form-pK}) and the value 
determined by the balance between the Hubble mass and the
nonrenormalizable term, is written as
\begin{equation}
    m_{\phi} \sim 2.4 \Omega_Q \beta_{\ell}^{1/4}
    m_{3/2,{\rm GeV}}^{-1/2} \ {\rm GeV},
\end{equation}
for $n=7$. We thus get the charge of the produced Q balls:
\begin{equation}
    Q \sim 1.9\times 10^{28} \Omega_Q^{-28/5}\beta_{\ell}^{-4/5}
    T_{RH,5}^4 m_{\phi,{\rm TeV}}^{36/5},
\end{equation}
where we plot for $T_{RH}=10^3$ and 80 GeV for the upper and the lower
thick dotted lines, respectively, in Fig.~\ref{fig18}. Taking into
account the condition $T_{eq} \gtrsim m_{\phi}$, we obtain
\begin{equation}
    Q \gtrsim 6.1\times 10^{29} \Omega_Q^{-32/5} \beta_{\ell}^{-1}
    m_{\phi,{\rm TeV}}^{12},
\end{equation}
where Eq.(\ref{Trh-m-phi}) is used. We plot this constraint by the
thick dashed line in Fig.~\ref{fig18}. The allowed region for the dark 
matter Q balls, which are also the source for the baryons in the
universe, is shown by thick solid line. This region does not suffer
from the current experimental limit and this dark matter Q ball with 
$Q \sim 10^{25}$ and $m_{3/2} \sim 100$ keV may be detected by TA in
the future. Notice that the reheating temperature is low enough to
avoid the cosmological gravitino problem. If we abandon to explain the
baryons in the universe by the charge evaporation from Q balls, the
dark matter Q balls may be found by TA and also OWL-AIRWATCH in the
future.

\subsection{Delayed Q balls in the generic logarithmic potential}
\begin{figure}[t!]
\centering
\hspace*{-7mm}
\leavevmode\epsfysize=10cm \epsfbox{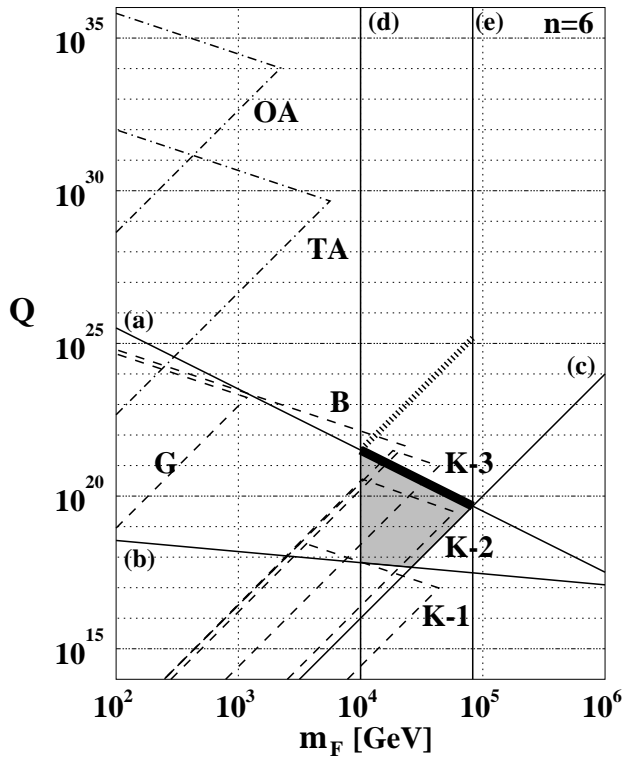}\\[2mm]
%\vspace{-3mm}
\caption[fig19]{\label{fig19} 
Summary of constraints on the parameter space $(Q,M_F)$ for the
delayed-formed Q balls with the generic logarithmic term domination
for $n=6$ and $m_{\phi}=100$ GeV. Experimental lines are the
same. Lines (a), (b), and (c) denote the B-DM, survival, and stability
conditions, respectively. Line (d) represents the condition 
$M_F\gtrsim T_{eq}$, and line (e) is just the upper limit for $M_F$.}
\end{figure}

\begin{figure}[t!]
\centering
\hspace*{-7mm}
\leavevmode\epsfysize=10cm \epsfbox{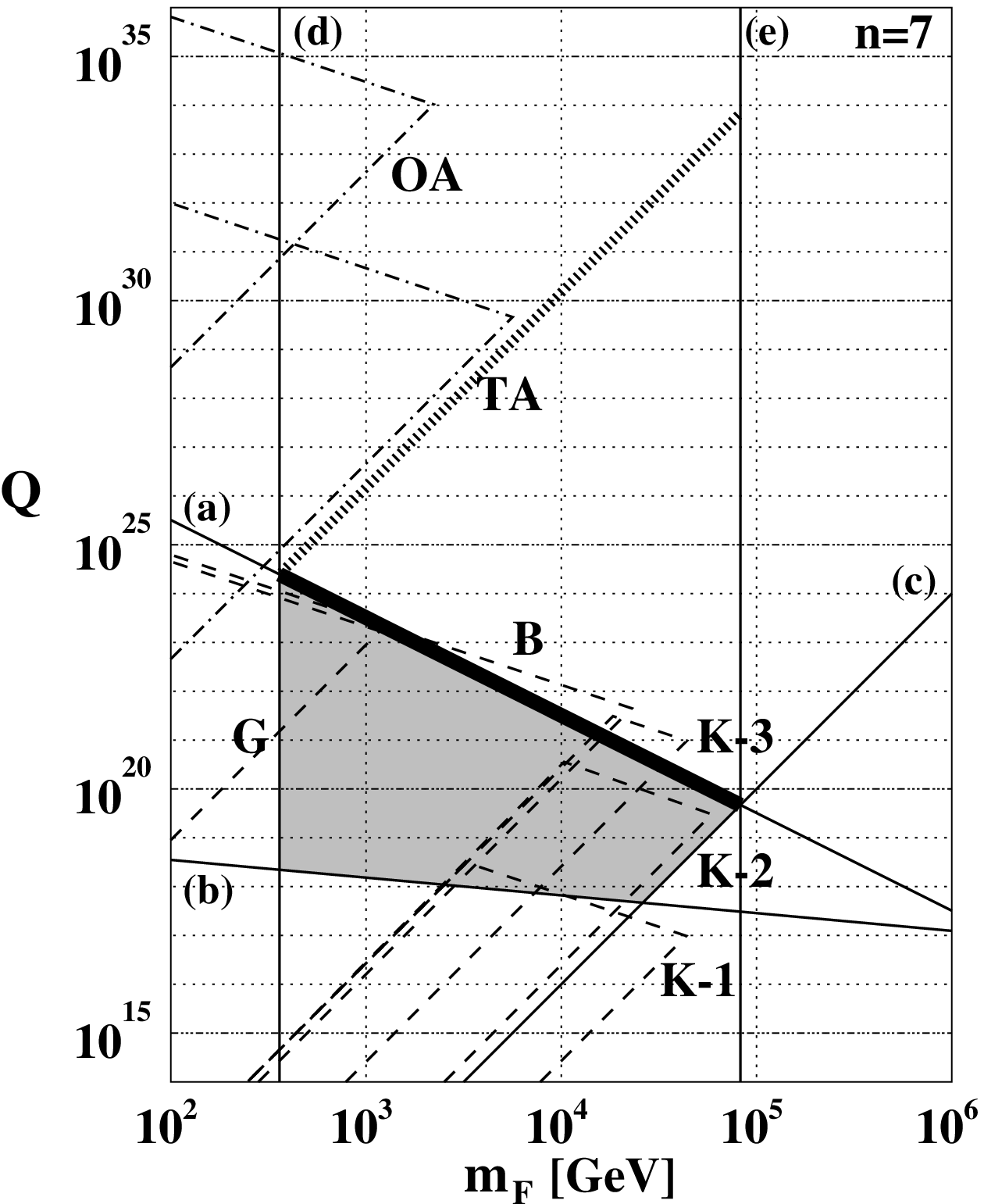}\\[2mm]
%\vspace{-3mm}
\caption[fig20]{\label{fig20} 
Summary of constraints on the parameter space $(Q,M_F)$ for the
delayed-formed Q balls with the generic logarithmic term domination
for $n=7$ and $m_{\phi}=100$ GeV. Experimental lines are the
same. Lines (a), (b), and (c) denote the B-DM, survival, and stability
conditions, respectively. Line (d) represents the condition 
$M_F\gtrsim T_{eq}$, and line (e) is just the upper limit for $M_F$.}
\end{figure}

\begin{figure}[t!]
\centering
\hspace*{-7mm}
\leavevmode\epsfysize=10cm \epsfbox{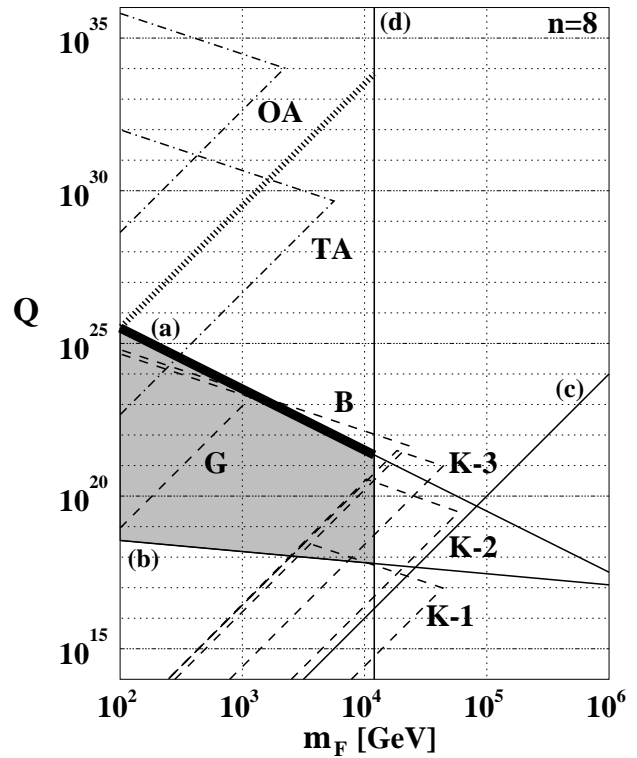}\\[2mm]
%\vspace{-3mm}
\caption[fig21]{\label{fig21} 
Summary of constraints on the parameter space $(Q,M_F)$ for the
delayed-formed Q balls with the generic logarithmic term domination
for $n=8$ and $m_{\phi}=100$ GeV. Experimental lines are the
same. Lines (a), (b), and (c) denote the B-DM, survival, and stability
conditions, respectively. Line (d) represents the condition 
$T_{RH} \gtrsim 10$ MeV.} 
\end{figure}

As mentioned in the previous section, we have obtained the consistent
scenarios of the dark matter Q ball explaining the baryons of the
universe naturally, with the initial conditions determined by the
balance of the Hubble and the nonrenormalizable terms in the effective 
potential, with $n=6$, 7, and 8. However, the parameter space is
restricted by the current experiments. We plot the allowed regions and 
experimentally excluded regions for $n=6$, 7, and 8 in
Figs.~\ref{fig19}, \ref{fig20}, and \ref{fig21}, respectively. In
these figures, lines (a), (b), and (c) are the B-DM, survival, and
stability conditions, respectively. Lines (d) in Figs.~\ref{fig19} and 
\ref{fig20} are the condition $M_F \gtrsim T_{eq}$, while 
$T_{RH} \gtrsim 10$ MeV is shown as line (d) in
Fig.~\ref{fig21}. Lines (e) in Figs.~\ref{fig19} and \ref{fig20} are
just the upper limit for $M_F$. 

We can see some parameter regions which are experimentally allowed in
each figures. Typically, there are two types: One is for 
$Q \sim 10^{20}$ and $M_F \sim 5\times 10^4$ GeV, and the other is for
$Q \sim 10^{25}$ and $M_F \sim 3\times 10^2$ GeV. In the latter case,
the dark matter Q balls may be detected in the future experiments. 
Notice that the required reheating temperature is rather too low 
(100 MeV $-$ 10 GeV), which is very difficult to realize in the
actual inflation model.

\section{Conclusion}
We have investigated thoroughly the Q-ball cosmology (the baryogenesis
and the dark matter) in the gauge-mediated SUSY breaking model. Taking 
into account thermal effects, the shape of the effective potential has 
to be altered somewhat, but most of the features of the Q-ball
formation derived at zero temperature can be applied to the finite
temperature case with appropriate rescalings. We have thus found that
Q balls are actually formed through Affleck-Dine mechanism in the
early universe. 

We have sought for the consistent scenario for the dark matter Q ball,
which also provides the baryon number of the universe
simultaneously. For the consistent scenario, we adopt the
nonrenormalizable superpotential in order to naturally give the
initial amplitude of the AD field and the source for the field
rotation due to the A-term. As opposed to our expectation, very narrow
parameter region could be useful for the scenario in the situations
that the Q balls are produced just after the baryon number
creation. In addition, we have seen that current experiments have
already excluded most of the successful parameter regions. 

Of course, if the A-terms and/or the initial amplitude of the AD field 
are determined by other mechanism, the cosmological Q-ball scenario
may work. Then, the stable dark matter Q balls supplying the baryons
play a crucial role in the universe. 

We have also found the new situations that the Q-ball formation takes
place when the amplitude of the fields becomes small enough to be in
the logarithmic terms in the potential, while the fields starts its
rotation at larger amplitudes where the effective potential is
dominated by the gravity-mediation term with {\it positive} K-term. 
This allows to produce Q balls with smaller charges while creating
larger baryon numbers. In this situation, there is wider allowed
regions for naturally consistent scenario, although the current
experiments exclude most of the parameter space. Notice also that
rather too low reheating temperature is necessary for larger $n$
scenario to work naturally. This aspect is good for evading the
cosmological gravitino problem, while it is difficult to construct the
actual inflation mechanism to get such low reheating temperatures, in
spite of the fact that the nucleosynthesis can take place successfully
for the reheating temperature higher than at least 10 MeV.

So far we have investigated the possibility that the dark matter Q 
balls which supply the baryon number of the universe could be really
produced in the naturally consistent scenario, where the initial
amplitude of the AD field and the A terms for the field rotation are
determined by the nonrenormalizable superpotential. Here, we have 
explicitly imposed that the produced Q balls must be stable against
the decay into nucleons in order for Q balls themselves to be the dark
matter of the universe. However, there may be such situations that Q
balls decay into lighter particles when the stability condition is
broken. This situation is almost the same as the usual AD
baryogenesis when we consider only the baryons in the
universe. Moreover, if the decay products include the lightest
supersymmetric particles (LSPs), they may account for the dark matter
of the universe. \footnote{
We thank M. Fujii for indicating the situation that the Q balls can
decay and become the source for the baryons and the LSP dark matter
simultaneously in the gauge-mediated SUSY breaking model. }

If this is the case, gravitino may be the LSP to be the dark
matter. The number density of LSP depends on the lifetime of the Q
ball, as investigated in Ref.~\cite{EnMc2} for the gravity-mediated
SUSY breaking model. If it decays before the temperature drops the
freeze-out temperature of LSPs, the density follows the thermal
value. On the other hand, when its lifetime is long enough, the ratio
of the number of the baryons to that of LSPs is fixed, and we can
directly estimate the relation between the amounts of the baryons and
the LSP dark matter of the universe.

\section*{Acknowledgments}
The authors are grateful to M. Fujii for useful discussion. M.K. is
supported in part by the Grant-in-Aid, Priority Area ``Supersymmetry
and Unified Theory of Elementary Particles''(No.~707). Part of the 
numerical calculations was carried out on VPP5000 at the Astronomical
Data Analysis Center of the National Astronomical Observatory, Japan.

\end{document}